\documentclass[preprint,authoryear,12pt]{elsarticle}

\usepackage{amssymb}

\usepackage[%
breaklinks=true,%
colorlinks=true,%
pdfauthor={Iglesias et al.},%
pdftitle={Ejectivity of a long duration active region}%
]{hyperref}

\usepackage{lineno}
\usepackage{graphicx,paralist,epstopdf,subfigure}
\usepackage{boldline}
\usepackage{lscape}
\usepackage{journals_abbreviations}
\usepackage{natbib}
\journal{Advances in Space Research}

\begin{document}


\begin{frontmatter}


\title{Analysis of a long-duration AR throughout five solar rotations: Magnetic properties and ejective events}

\author[a,b]{Francisco A. Iglesias \corref{mycorrespondingauthor}}
\cortext[mycorrespondingauthor]{Corresponding author}
\ead{franciscoiglesias@frm.utn.edu.ar}

\author[a,b]{Hebe Cremades}
\ead{hebe.cremades@frm.utn.edu.ar}
\author[a]{Luciano A. Merenda}
\ead{lucianomerenda3@gmail.com}
\author[c,d]{Cristina H. Mandrini}
\ead{mandrini@iafe.uba.ar}
\author[e]{Fernando M. L{\'o}pez}
\ead{fmlopez@conicet.gov.ar}
\author[c]{Marcelo C. L{\'o}pez Fuentes}
\ead{lopezf@iafe.uba.ar}
\author[f]{Ignacio Ugarte-Urra}
\ead{ignacio.ugarte-urra@nrl.navy.mil}

\address[a]{Universidad Tecnol{\'o}gica Nacional\,--\, Facultad Regional Mendoza, CEDS, Rodr\'iguez 243, 5500, Mendoza, 
Argentina}
\address[b]{Consejo Nacional de Investigaciones Cient{\'i}ficas y T{\'e}cnicas (CONICET), Godoy Cruz 2290, C1425FQB, 
Buenos Aires, Argentina}
\address[c]{Instituto de Astronom{\'i}a y F{\'i}sica del Espacio (IAFE, UBA-CONICET),  CC. 67, Suc. 28, Buenos Aires, 
1428, Argentina}
\address[d]{Facultad de Ciencias Exactas y Naturales (FCEN), Universidad de Buenos Aires (UBA), Intendente Guiraldes 
2160, C1428EGA, Buenos Aires, Argentina}
\address[e]{Instituto de Ciencias Astron{\'o}micas, de la Tierra y del Espacio (ICATE), CONICET-UNSJ, Av. España 1512 
sur, 5400, San Juan, Argentina}
\address[f]{Space Science Division, Naval Research Laboratory, 4555 Overlook Ave SW, Washington DC, 20375, USA}

\begin{abstract}
	Coronal mass ejections (CMEs), which are among the most magnificent solar eruptions, are a major driver of space 
	weather and can thus 
	affect diverse human technologies. Different processes have been proposed to explain the initiation and release of 
	CMEs from solar active regions (ARs), without reaching consensus on which is the predominant scenario, and thus 
	rendering impossible to accurately predict when a CME is going to erupt from a given AR. To investigate AR magnetic 
	properties that favor CMEs production, we employ multi-spacecraft data to analyze a long duration AR (NOAA 11089, 
	11100, 11106, 11112 and 11121) throughout its complete lifetime, spanning five Carrington rotations from July to 
	November 2010. We use data from the Solar Dynamics Observatory to study the evolution of the AR magnetic 
	properties during the 
	five near-side passages, and a proxy to follow the magnetic flux changes when no magnetograms are available, i.e. 
	during far-side transits. The ejectivity is studied by characterizing the angular widths, speeds and masses of 108 
	CMEs that we associated to the AR, when examining a 124-day period. Such an ejectivity tracking was 
	possible thanks to the mulit-viewpoint images provided by the Solar-Terrestrial Relations Observatory and Solar 
	and Heliospheric Observatory in a quasi-quadrature 
	configuration. We also inspected the X-ray flares registered by the GOES satellite and found 162 to be 
	associated to the 
	AR under study. Given the substantial number of ejections studied, we use a statistical approach instead of a 
	single-event 
	analysis. We found three well defined periods of very high CMEs activity and two periods with no mass 
	ejections that are preceded or accompanied by characteristic changes in the AR magnetic flux, free magnetic energy 
	and/or presence of electric currents. Our large sample of CMEs and long term study of a single AR, provide further 
	evidence relating AR magnetic activity to CME and Flare production.
    
\end{abstract}


\end{frontmatter}

\parindent=0.5 cm


\section{Introduction}
\label{sec:intro}

Active regions (ARs) are areas of intense magnetic field 
concentration on the Sun that 
are constantly evolving throughout their lifetime, typically ranging from days to a few moths (see e.g. 
\citealt{vandriel2015} and references therein). From their generation, linked to the emergence and concentration of 
new photospheric magnetic flux, to their decay, partially driven by the spatial spreading and cancellation of such 
flux, ARs are centers of diverse magnetic activity. They provide vital constraints to model the underlying dynamo 
process \citep{vandriel2015} and are also the main source region of different kinds of transient phenomena, such as 
solar 
flares (see e.g. \citealt{Priest2002}) and coronal mass ejections (CMEs). CMEs 
involve the 
fast release of large amounts of mass and 
magnetic field from the solar corona into the interplanetary medium (sometimes exceeding $2500$ kms$^{-1}$ and 
$10^{16}$ g), see e.g. 
\citealt{Webb2012}. They 
produce 
significant perturbations in the solar wind and can strongly influence the geomagnetic environment conditions, a.k.a 
space weather, see e.g. \cite{bothmer2007} and \cite{zhang2018}.

Magnetic energy dominates other forms of energy in the low corona, particularly near ARs, where magnetic 
pressure overcomes plasma pressure and drives the matter dynamics. The occurrence of a CME is then of magnetic nature, 
as summarized in \cite{Green2018} requiring (a) the previous build-up of free magnetic energy stored in the 
non-potential 
core field, which may or may not contain a filament and is generally located above the polarity inversion line (PIL) of 
ARs; (b) a destabilizing mechanism that 
triggers the eruption of the core field; and (c) a driving mechanism that powers the 
ejection of the core field from the low to the high corona while interacting with the overlying strapping field.

Several mechanisms contribute to build up 
non-potential energy and magnetic helicity in the coronal field associated to ARs. These include, among others, sunspot 
rotation, the 
frequent 
emergence of twisted magnetic flux tubes (or 
flux ropes, see e.g. \citealt{hood2009,poisson2015a}) and the 
stress 
produced in the field lines by shearing photospheric flows (e.g. \citealp{mactaggart2010}). There is 
substantial observational evidence 
of the presence in the solar atmosphere of the topological features (e.g. S-shaped loops, magnetic tongues, etc.) and 
electric currents 
associated to such a non-potential field, e.g. \cite{Rust-Kumar1996,mckenzie2008,koleva2012, 
Jiang2014, poisson2015b}. Abrupt magnetic reconfigurations, 
associated to the reconnection of field lines, transform large 
amounts of the free magnetic energy stored in the coronal field into kinetic and thermal energy, powering eruptive 
events 
such as CMEs and flares, e.g. 
\citep{Kliem2014,Aulanier-etal2010}.

CMEs are commonly associated with flux rope eruptions, e.g. \cite{li2012,Jiang2014,Vourlidas-etal2013}. 
Different 
 mechanisms have been proposed and evidenced in the literature to explain CME triggering, including flux emergence 
 (e.g. 
\citealt{Chen-etal1997} and \citealt{Manchester-etal2004}), reconnection of field lines below (tether-cutting model, 
\citealt{moore1992}) or above (breakout model, \citealt{Antiochos-etal1999}) the flux rope, excess of twist in the flux 
rope (kink instability, \citealt{Torok-Kliem2005}) and others (e.g. \citealt{Amari-etal2000,Lin-etal2004} and 
\citealt{Aulanier-etal2010}).

Once the core field is destabilized, it rises stretching and pushing aside the overlying coronal 
field. It can be the case that this	strapping field restrains the rising core field, preventing its ejection and 
producing a 
confined CME, see e.g. \cite{Torok-Kliem2005} and \cite{moore2001}. In any case, the above-named trigger mechanisms are 
not able to explain the observed acceleration and expansion of CMEs in the low corona. Instead, two driving 
processes have been proposed, namely the torus instability \cite[a.k.a. flux-rope catastrophe model, e.g.][]{Kliem2006, 
Aulanier-etal2010}, which occurs 
when the outward 
magnetic pressure of 
the flux rope exceeds the inward magnetic tension provided by the external field; and the 
flare-reconnection \cite[e.g.][]{forbes2018}, that describes the successive magnetic 
reconnections occurring at the 
vertical current sheet formed 
below the rising core field, and its associated flaring activity.

Mainly due to the lack of routine magnetic field measurements of the corona, no clear consensus has been 
reached 
regarding which of the named trigger and driving mechanisms, or what 
combination of them, is the predominant, see e.g. the discussions in \cite{Webb2012,Green2018}. Moreover, the activity 
of ARs varies during their lifetime. Flares are common in the emergence and stable phase, decreasing 
in number with the reduction of flux density during decay. On the other hand, CME production is generally 
low during the emergence of young ARs, however, it can persist or even increase 
during the stable and decay phases, see Sect. \ref{sec:discussion} and e.g. \cite{li2012,demoulin2002}. Because of 
this, 
studying 
the evolution of the magnetic properties of ARs in connection with their 
associated eruptive events is an active area of research. The vast literature includes short-term (a fraction of the AR 
lifetime), detailed analyses focusing on, e.g. comparative CME-production (e.g. \citealt{Cremades-etal2015b, 
Murray2018}), pre- and 
post-eruptive coronal magnetic field topology (e.g. \citealt{Mandrini-etal2014,Mandrini2006,Chandra2011,Chandra2017}) 
and magnetic helicity 
evolution (e.g. \citealt{romano2014, demoulin2002,Mandrini2004b}). There are also investigations of the long-term (time 
scales covering a full AR 
lifetime or more) 
evolution of, e.g the magnetic influence of AR plasma flows (e.g. \citealt{harra2017, zangrilli2016} and 
\citealt{ko2016}), 
the CME production rate along the solar cycle (e.g. \citealt{gopalswamy2003} and \citealt{riley2006}), the global 
magnetic 
field and its associated CME production \citep{petrie2013}, and the continuous tracking of some AR magnetic 
properties (e.g. \citealt{demoulin2002} and \citealt{green2002}), among many others, see e.g. the review by 
\cite{vandriel2015}.

Given the present fleet of Sun-observing missions, up to date we can only obtain magnetograms of the portion of the 
solar surface that is facing Earth, i.e. the near side. Moreover, limb darkening and spherical effects harm the quality 
of the magnetograms obtained from a fixed Earth perspective, e.g. the noise properties of the tangential and radial 
field components change from disk center to the limb. Therefore, all long-term studies cited above were either done 
on ARs that live less than approximately half a solar rotation, are restricted to only the intervals where the AR is on 
the near side, or have used a proxy to estimate magnetic properties when the AR is on the far side, such as using 
304\,\AA~intensity images or constrained magnetic surface flux transport models to estimate total flux, see e.g. 
\cite{ugarteurra2015}. On the contrary, the above-named limiting factor is not present when studying the CME 
production of an AR. There is the possibility of continuous tracking of the CME production of an AR using a combination 
Sun-observing spacecraft such as SDO (Solar Dynamics Observatory; \citealt{Pesnell-etal2012}) and/or SOHO (Solar and 
Heliospheric Observatory; \citealt{Domingo-etal1995}) plus the two STEREO (Solar-Terrestrial Relations Observatory; 
\citealt{Kaiser-etal2008}), provided 
that the latter are favorably located so as to track the AR during its far side passage, i.e. nearly in 
quadrature with the Sun-Earth line ($\approx180^\circ$ apart). This combination of observatories offers a unique 
opportunity to examine the CME production continuously during one or more full solar rotations.

The present work reports on the CME and X-ray flare production of a long duration AR (NOAA 11089, 11100, 
11106, 11112 and 11121) 
throughout its complete lifetime, spanning five Carrington rotations (CRs) from July to November 2010. We also analyze 
the 
evolution of some of the AR photospheric magnetic properties (magnetic flux, current helicity and a proxy of the 
photospheric free magnetic energy, see Sect. \ref{sect:dataAR} for exact definitions) to study their 
relationship with the frequency and 
properties of the 
ensued CMEs. Given the substantial number of mass ejections studied (108) and their clustering in bursts, we do not 
focus on single events but relate 
the long-term (few days) variation of the AR 
magnetic properties to the occurrence of bursts of CMEs, i.e. high CME activity periods. The rest of this work is 
organized as follows. Section \ref{sec:data} presents the methodology and 
analyzed data, including that of the AR (Sect. \ref{sect:dataAR}) acquired during its near-side (using  HMI and 
AIA\footnote{The Helioseismic and Magnetic Imager (HMI, \citealt{scherrer2012}) and the Atmospheric Imaging Assembly (AIA, 
\citealt{Lemen-etal2012}) are both onboard the SDO spacecraft (in geosynchronous orbit).} onboard SDO, and 
MDI\footnote{The Michelson Doppler 
Imager (MDI, 
\citealt{scherrer1995}) onboard the SOHO spacecraft (located at Lagrangian point 1 of the Sun-Earth system).} onboard 
SOHO) and far-side (using SECCHI EUVIs\footnote{The 
Extreme--Ultraviolet 
Imagers (EUVIs) are part of the  Sun--Earth Connection Coronal and Heliospheric Investigation experiment (SECCHI,  
\citealt{Howard-etal2008}) onboard of the two STEREO spacecrafts (orbiting the sun in opposite directions).}  onboard 
STEREO) transits. 
Sect. 
\ref{sect:dataCME} describes the SECCHI 
and LASCO\footnote{The Large-Angle and Spectrometric Coronagraph Experiment (LASCO, \citealt{Brueckner-etal1995}) 
onboard 
the SOHO spacecraft.} data that allowed us to track the AR, identify its associated CMEs and derive 
their main properties. Sect. \ref{sect:dataFlares} introduces the Geostationary Operational Environmental 
Satellite (GOES) data used to identify X-ray flares 
originating in the AR. Sect. \ref{sec:results} presents and describes the resulting time series that drive the 
discussion and conclusions given in 
Sect. \ref{sec:discussion}.

\section{Data sets and methodology}
\label{sec:data}
After Solar Cycle 23, a long solar minimum of over two years, and more than 800 days without 
sunspots, a 
series of long-duration ARs emerged on the Sun. Many of these had strong magnetic activity with flares, filament 
eruptions and CMEs, see e.g. \cite{Schrijver2011,Liu2012a, li2012, Mandrini-etal2014}.

The inspected AR was born in the far side of the Sun and appeared for the first time on the east limb 
as NOAA AR 11089 on 19 July 2010, persisting for approximately five CRs until mid-November 2010. During that period, the STEREO twin spacecraft were approaching a quadrature configuration with 
respect to Earth, i.e. they were $\approx148^\circ$ and $\approx168^\circ$ apart at the beginning and end of the 
mentioned time interval, 
respectively. At the same time, the SDO mission was beginning its operational phase, providing views of the AR from 
Earth's perspective. 
Numerous episodes of flux emergence and ejective activity were observed during the lifetime of the investigated AR. As a 
consequence, it has been subject of independent, short-term studies that address different aspects and stages of its 
evolution, e.g. \cite{guo2013,zuccarello2014,Mandrini-etal2014,Cremades-etal2015b}. In the latter two articles, AR 
11121 is 
analyzed together with the closely related AR 11123, which emerged within AR 11121 during November 2010. 

\begin{landscape}
	\begin{figure}
		\raggedright
		\begin{subfigure}{}	
			\raisebox{0.78cm}{\includegraphics[width=3.5in]{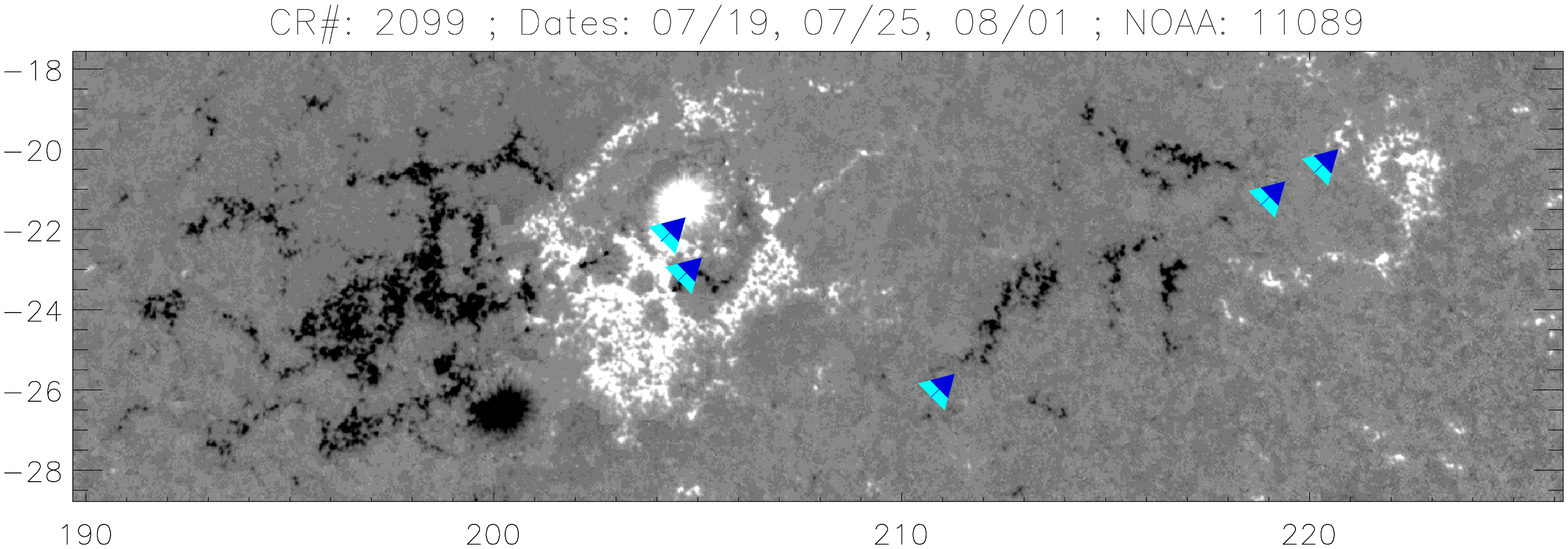}}
		\end{subfigure}
		\begin{subfigure}{}	
			\raisebox{0cm}{\includegraphics[width=3.5in]{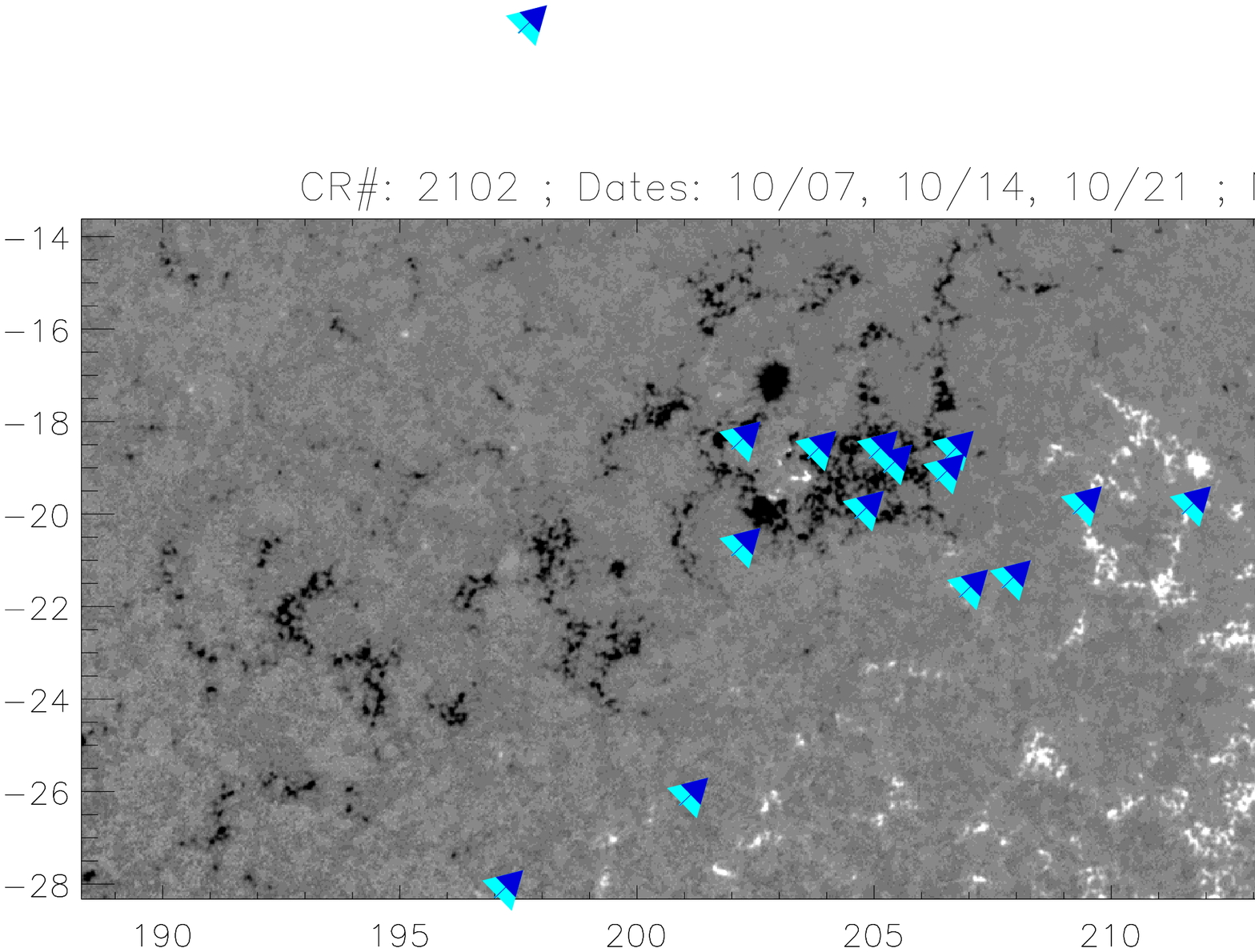}}
		\end{subfigure}
		\begin{subfigure}{}	
			\raisebox{0cm}{\includegraphics[width=3.5in]{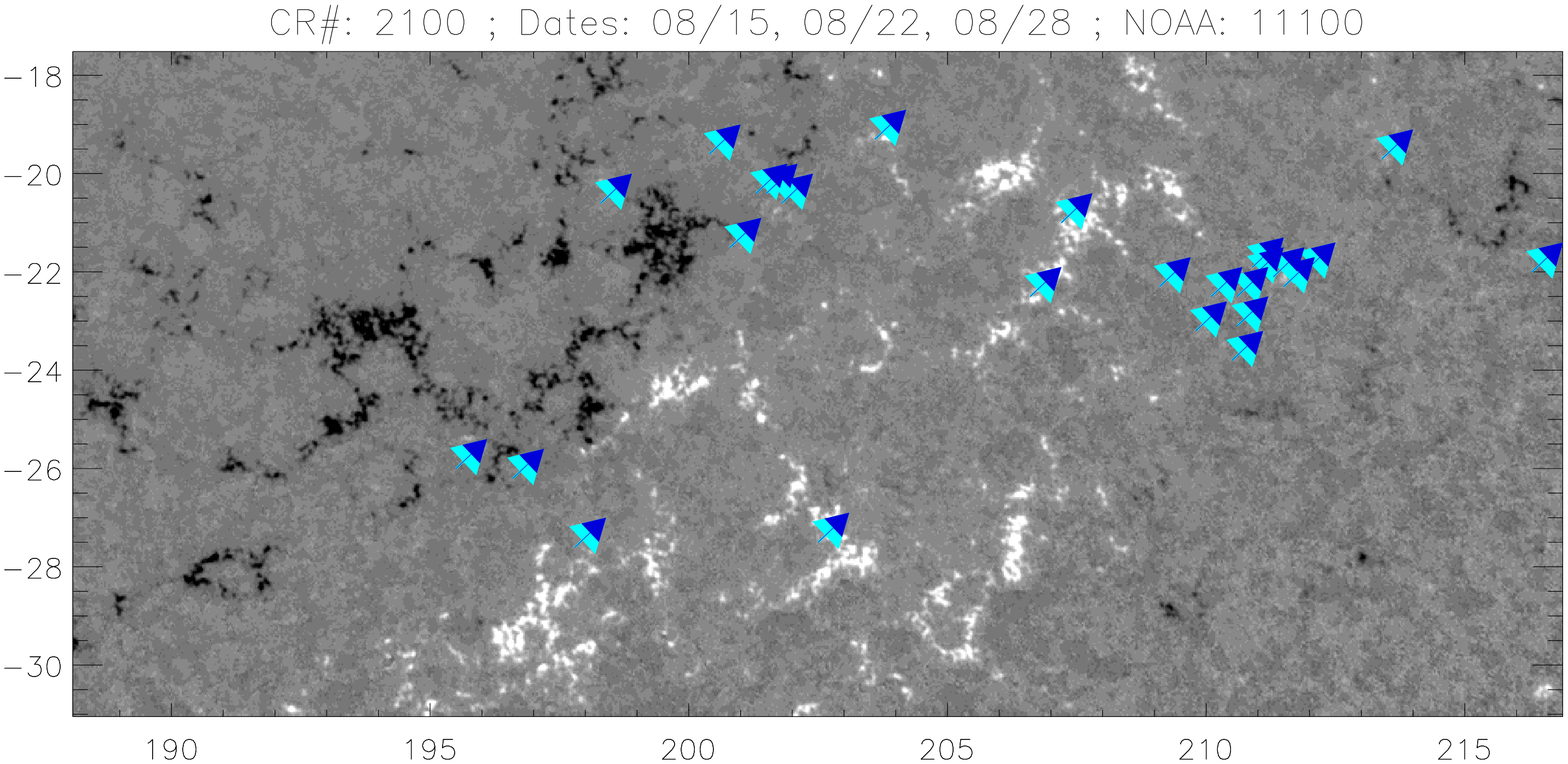}}
		\end{subfigure}
		\begin{subfigure}{}	
			\raisebox{0cm}{\includegraphics[width=3.5in]{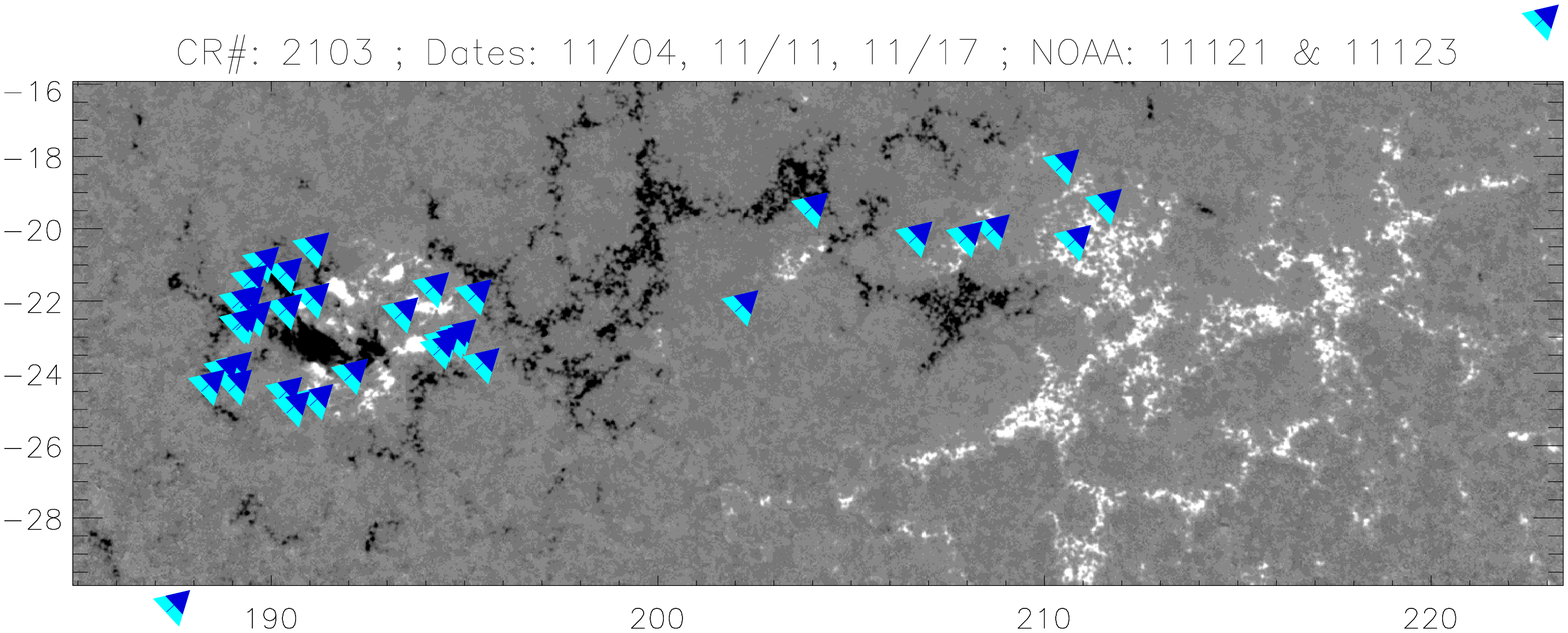}}
		\end{subfigure}
		\begin{subfigure}{}	
			\includegraphics[width=3.5in]{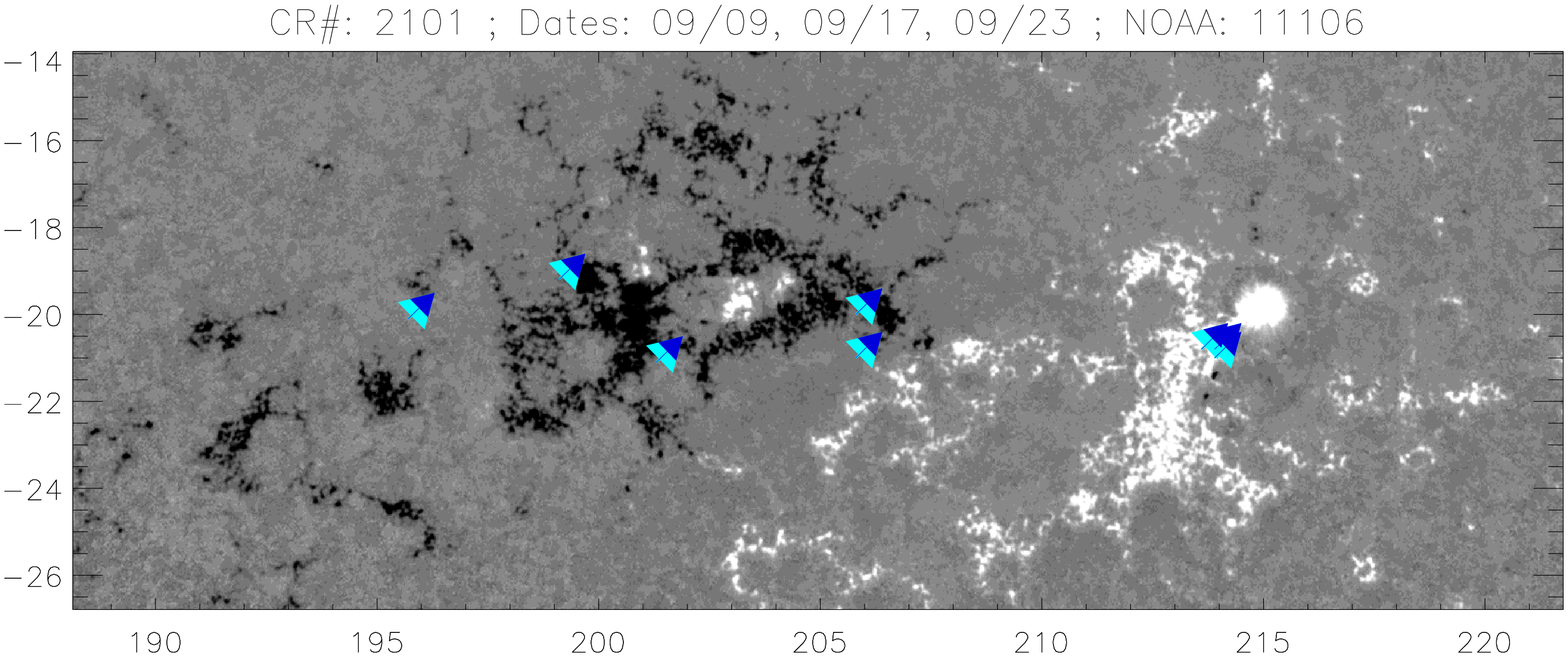}
		\end{subfigure}	
		\caption{SHARP patches showing line-of-sight magnetograms of the investigated AR during its five 
		central-meridian passages. The spatial scales are in degrees. The color scale ranges from -500 to 500 G with 
		fields pointing outwards of the solar surface in white. The title of each patch gives the CR number; the dates 
		(in the format month/day) when the region was at the west limb, central meridian and east limb; and the 
		assigned NOAA number, respectively. The blue arrows point to the approximate location of EUV 
		brightenings or 
		filamentary eruptions associated to 96 CMEs that occurred at different times within each solar rotation, see 
		Sect. \ref{sect:dataCME} for extra details.}
		\label{fig:armag}
	\end{figure}
\end{landscape}

The photospheric imprints of the inspected AR can be seen in Fig. \ref{fig:armag}, which presents the line-of-sight 
magnetograms during its five central meridian passages. We also display the different NOAA numbers that were assigned 
to the AR after each new solar rotation, together with the dates at which it appeared on the east limb, was at central 
meridian, and disappeared on the west limb.  The AR is seen to constantly evolve, starting with a configuration 
	predominantly formed by two bipoles in CR 2099 (top-left panel in Fig. \ref{fig:armag}). During the near-side transit 
	of the AR the bipoles do not present strong photospheric interaction, e.g. cancellation of opposite polarities , 
	and the 
	western, weaker bipole diffuses to be absent in CR2100. Moreover, we did not find obvious signs of coronal 
	magnetic interaction (in terms of simultaneous EUV 
	brightenings occurring over both bipoles) and the CME activity was low (as is frequent in young ARs, see Sect. 
	\ref{sec:intro}). After 
	this, the AR adopts a predominantly bipolar configuration 
	from CR 2100 to CR 2102. During the last rotation (CR 2103, shown in the bottom-right panel of Fig. \ref{fig:armag}) 
	the 
	bipolar AR 
	11123 emerged in the negative polarity area 
	of the decaying AR 11121, strongly incrementing the CME activity (see Sect. \ref{sec:discussion}). 

In the following three subsections we 
describe the data sets 
employed to study the evolution of the photospheric magnetic field,  as well as the CME and X-ray flare 
production 
along the AR lifetime. 

\subsection{Photospheric properties}
\label{sect:dataAR}

Magnetic flux evolution is expected to show correspondence with eruptive activity given that photospheric motions,  
including those related to flux emergence, have been 
pointed out as a possible trigger of CME eruptions, e.g. \cite{chen2011,vandriel2015}. AR parameters 
characterizing the field non-potentiality, e.g. electric currents or free magnetic energy, have been also widely 
accepted to be related to solar eruptions, e.g. \cite{Canfield-etal1999,falconer2006,wang2008,guo2013,liu2016}. 
Here we analyze temporal series of the total unsigned magnetic flux ($\sum |B_z| dA$), mean current helicity ($\bar{Hc} 
\propto \frac{1}{N}\sum B_z J_z$), total unsigned current helicity ($Hc\propto \sum |B_z J_z|$), 
and 
a proxy of the mean photospheric free magnetic energy density ($\frac{1}{N}\sum 
(\mathbf{B}^{obs}-\mathbf{B}^{pot})^2$), where $\mathbf{B}$ is the magnetic field vector with radial component $B_z$, 
$N$ is the number of spatial resolution elements (magnetogram pixels) contributing to the computation, $J_z$ is the 
current density in the radial direction and $dA$ gives the surface area covered by each pixel. In the free energy 
density expression, $\mathbf{B}^{obs}$ and 
$\mathbf{B}^{pot}$ represent the observed and potential field (modeled) respectively. These data series are obtained 
from the 
Space-weather HMI 
Active Region Patches (SHARPs; \citealt{bobra2014}) during the AR near-side passages, covering 124 days 
(from 17 July to 18 November 2010).  The SHARPs (see Fig. \ref{fig:armag}) are standard data products extracted from HMI 
12\,min-cadence, full-disk, vector magnetograms. Only pixels with transverse field strength that exceed the 
azimuth disambiguation noise threshold ($\approx150$\,G) contribute to the patch parameters, see \cite{bobra2014} for 
extra details.

We additionally estimated the time evolution of the total unsigned magnetic flux in the AR using two other independent 
sources. The 
first one are full-disk, level-1.8 MDI magnetograms. These data are the average of 5 line-of-sight magnetograms with a 
cadence of 30\,s and a noise error of 20\,G per pixel. They are constructed once every 96 min and have an error in 
the flux density per pixel 
of 9\,G. Following \cite{Green2003}, to find the AR flux, a polygonal contour is fitted around the AR. The shape of the 
contour is given by the large gradient between the AR  and  the  network  fields.  Within  this  region,  the flux is 
calculated for all pixels with absolute fields above 50\,G. Due to the limitations related to the projection effects 
involved in the magnetic flux determination, we only show values corresponding to the dates when the AR was on central 
meridian $\pm$\,4 days. The resulting MDI flux curves were divided by a 
cross-calibration factor of 1.40 to make them comparable to the HMI results, see \cite{Liu2012b, Chertok2019}. 

The second method employed, estimates the total unsigned magnetic flux using its strong positive correlation with the
total brightness of the He II 304\,\AA~spectral line (e.g. \citealt{Schrijver1987}). We used the technique developed
by \cite{ugarteurra2015,ugarte-urra2018}, which allows estimating the flux when magnetograms are not available, by 
employing
intensity-only 304\,\AA~images as a proxy. In this technique, a synchronic Carrington map covering most of the 
solar surface is assembled by combining quasi-simultaneous images of the 304\,\AA\,channels of the STEREO EUVIs and SDO 
AIA instruments. The covered solar surface depends on the SDO and STEREO spacecraft locations. The EUVI 
and
AIA data are previously corrected to account for limb darkening effects and the time-dependent, cross-calibration of
the instruments. The EUV synchronic maps allow tracking the movement of the AR on the solar surface, in our case at a
cadence of one image every 6h. A fixed-sized, square box is defined around the AR of interest in each map, and the 
total photon
flux is computed by adding the contributions of all pixels within the box that also belong to the AR. Pixels with a flux
two standard deviations above the normal distribution of quiet Sun flux in full maps are considered to belong to an AR.
The total photon flux can then be translated to the total unsigned magnetic flux using a known power-law
relationship, see \cite{ugarte-urra2018,ugarteurra2015} for extra details. This technique can be applied under
complex scenarios, e.g. strong emergence within an AR as is the case of ARs 11121 and 11123. The 304-proxy
method may, however, underestimate the magnetic flux when sunspots are present within the considered area.
Sunspots, while contributing significant magnetic flux to active regions, do not emit strongly in coronal or
chromospheric spectral lines such as those contained in the 304\,\AA~bandpass. This caveat is already implicit in the
power-law relationship which was optimized for magnetic flux densities in the range 90--900G \citep{fludra2008}.
Note that, the differences between MDI and HMI spatial resolution plus the different lower-field thresholds employed
in the flux estimations named above, may introduce discrepancies between their results.

\subsection{CME productivity}
\label{sect:dataCME}

Using quasi-simultaneous observations from multiple vantage points,
we have tracked the AR in EUV wavelengths for five solar rotations and identified its productivity in terms of 
white-light ejecta. The nearly-quadrature observations enabled a better estimation of the ejecta source regions, a 
sometimes challenging task, in particular for events propagating along the Sun-observer line (halo CMEs) which can lack 
structure, be diffuse and dim (see e.g. \citealt{Lara-etal2006, Cremades-etal2015b}). The location of the AR was 
decisive to determine which instrument was best to observe the region and its associated eruptive phenomena. 
During the 
AR near-side passages we examined it with AIA and HMI onboard SDO, while we detected the ensuing CMEs from a quadrature 
perspective using the STEREO/SECCHI COR2 coronagraph. Likewise, when the AR was close to the solar limbs (from Earth's 
viewpoint), we used the SECCHI EUVIs to track the region activity (EUVI-B for the east limb and EUVI-A for the west 
limb), and SOHO/LASCO C2 to identify the associated mass ejections. As the AR transited the far side of the Sun, we also 
used the SECCHI EUVIs to monitor its behavior, while the SECCHI COR2 coronagraphs were used to detect the associated 
eruptions. Note that we do not use LASCO C3 or
SECCHI HI data, to reduce the effects of the surrounding corona on the
derived CME properties, for instance due to solar wind acceleration or mass
loss.

After careful inspection of these observations, we compiled 108 CMEs that could be associated to the AR of 
interest, these are presented in Table \ref{tab:cme}. During this selection, we included all kinds of 
white-light ejecta that were 
discernible from the background corona in 
running-difference images, and at least in two consecutive images. For each CME we obtained the following properties,  
which are directly associated to the energy involved in the event given that generally wider CMEs are more massive 
and faster (see e.g. \citealt{gopalswamy2010}):

\begin{itemize}
	\item \textit{EUV brightening location and time}: We inspected regular and running difference 
	images of 
	the 171, 195 and 304\,\AA~channels to track back the initiation of the identified CMEs using their 
	maximum cadence ($\approx$10 min). We adopted the location of the maximum brightening 
	observed at any of these wavelengths, preferentially 304\,\AA.  Whenever we could identify an extended structure, 
	such as a filament as the 
	responsible for a given CME, we took instead the coordinates of the central point of that structure. This was done 
	to unambiguously associate the CMEs with the AR under study. 
	Similarly, the ejection time is defined as the 
	time of the first EUV brightening or filament eruption. Table \ref{tab:cme} lists these ejection times, along 
	with the time of the first appearance of the associated CME leading edge in white-light images of LASCO C2 or 
	COR2. 
	\item \textit{Angular width }(AW): The AW was measured in the set of coronagraphic images where the CME propagation 
	direction was closer to the corresponding plane of the sky \citep[e.g.][]{Cremades-etal2015b}. For completeness, we 
	also consulted the LASCO CME 
	Catalog 
	\citep{Yashiro-etal2004} and the Computer-Aided CME Tracking catalog (CACTus, \citealt{Robbrecht-Berghmans2004}).
	\item\textit{ Mass:} The coronal electron density can be estimated from the total brightness of white-light 
	coronagraphic images using the method introduced by \cite{Vourlidas2010b}, which makes use of the Thomson 
	scattering properties \citep{Billings1966} and assumes the electrons propagate predominantly in the plane 
	perpendicular to the line of sight. After assuming a given plasma composition, e.g. a mixture of completely ionized 
	hydrogen and 10$\%$ helium, the electron density can be translated to total mass, see e.g. \cite{Colaninno2009}. 
	Given the large number of events, we followed the implementation by \cite{Lopez2017} which uses 
	data from a single coronagraph  to derive the total mass of each event, by adding the 
	contribution of the mass associated to each pixel that belongs to the CME. The boundary of the CME is manually 
	selected by defining a freehand region of interest.
	\item\textit{Linear propagation speed:} Given that many of the selected events are not cataloged (44$\%$), we 
	derived their 
	plane-of-sky propagation speed from a linear fit to 
	height-time data points. These data points were obtained from a manual tracking of the CME leading-edge, 
	using  running-difference,	coronagraphic 
	images of LASCO C2 or SECCHI COR2 (depending on whether the direction of propagation was closer to one or the 
	other, see below). Therefore, for accelerated or decelerated events, the speed we derived represents their mean speed in 
	the 1.5-2 to 6 solar radii height range. The exact measurement interval within this range is event-dependent, e.g. 
	dim CMEs could not be generally followed across the complete field of view.
\end{itemize}

\begin{landscape}
\begin{table}[tp]
	\begin{center}
		\caption{Identification numbers (ID), dates (in dd/mm format) and times of 108 
		white-light ejective events (CMEs), originating from the AR under study. The times in columns 
		labeled EUV correspond to the 
		first brightening observed in AIA or EUVI EUV images. The times in columns labeled WL correspond to the first 
		appearance 
		in LASCO C2 or COR2 images. We also include the class of the associated GOES X-ray flare for 33 CMEs, as a 
		super index of the ID number. The horizontal lines delimit the beginning (thick lines) and end (thin lines) of 
		the high CME activity periods, see Sect. \ref{sect:dataCME}.}
		\label{tab:cme}
		{\small			
	\begin{tabular}{|llll|llll|llll|llll|}
		\hline	
		\textbf{ID} & \textbf{Date} & \textbf{EUV} & \textbf{WL} & \textbf{ID} & \textbf{Date} & \textbf{EUV} & 
		\textbf{WL} & \textbf{ID} & \textbf{Date} & \textbf{EUV} & \textbf{WL} & \textbf{ID}  & \textbf{Date} & 
		\textbf{EUV} & \textbf{WL} \\ \hline
		1  & 19/07 & 09:16 & 10:06 & 28 & 01/09 & 09:45 & 10:24 & 55$^{C}$ & 18/10 & 16:30 & 17:12 & 82$^{B}$  & 10/11 
		& 
		16:45 & 17:24 \\
		2  & 20/07 & 06:01 & 07:54 & 29 & 01/09 & 12:00 & 12:36 & 56$^{C}$ & 19/10 & 07:00 & 07:12 & 83$^{C}$  & 11/11 
		& 
		02:15 & 02:54 \\
		3$^{B}$  & 20/07 & 09:31 & 10:30 & 30 & 01/09 & 13:30 & 13:54 & 57$^{B}$ & 19/10 & 12:45 & 14:36 & 84  & 11/11 
		& 05:15 & 05:39 \\
		4$^{B}$  & 23/07 & 12:00 & 13:39 & 31 & 01/09 & 15:10 & 15:24 & 58 & 19/10 & 16:10 & 16:36 & 85$^{C}$  & 11/11 
		& 
		07:30 & 07:54 \\ 
		5  & 02/08 & 16:30 & 17:54 & 32 & 01/09 & 19:15 & 20:08 & 59 & 19/10 & 18:45 & 19:00 & 86$^{B}$  & 11/11 & 
		10:15 & 10:39 \\
		6  & 05/08 & 00:30 & 01:24 & 33 & 01/09 & 21:45 & 22:08 & 60 & 19/10 & 23:10 & 23:27 & 87$^{C}$  & 11/11 & 
		13:15 & 13:39 \\
		7  & 08/08 & 00:05 & 08:54 & 34 & 02/09 & 09:10 & 09:39 & 61$^{B}$ & 20/10 & 01:15 & 01:25 & 88$^{C}$  & 11/11 
		& 
		16:15 & 16:39 \\
		8  & 10/08 & 04:35 & 05:54 & 35 & 02/09 & 15:30 & 16:08 & 62$^{C}$ & 20/10 & 11:55 & 12:12 & 89$^{C}$  & 12/11 
		& 
		19:30 & 19:54 \\
		9  & 15/08 & 05:50 & 06:24 & 36 & 02/09 & 12:00 & 18:54 & 63 & 21/10 & 23:30 & 23:48 & 90$^{B}$  & 12/11 & 
		00:15 & 00:24 \\
		10 & 18/08 & 01:30 & 03:24 & 37 & 03/09 & 12:45 & 14:54 & 64 & 22/10 & 01:00 & 01:48 & 91$^{C}$  & 12/11 & 
		01:45 & 01:54 \\
		11 & 21/08 & 14:45 & 15:39 & 38 & 04/09 & 08:10 & 08:24 & 65 & 22/10 & 05:00 & 05:12 & 92$^{C}$  & 12/11 & 
		04:00 & 04:24 \\ \cline{5-8}
		12 & 23/08 & 12:00 & 23:24 & 39 & 05/09 & 07:45 & 08:24 & 66 & 23/10 & 01:00 & 01:58 & 93$^{C}$  & 12/11 & 
		08:10 & 08:24 \\
		13 & 27/08 & 11:30 & 10:12 & 40 & 06/09 & 00:05 & 00:24 & 67 & 23/10 & 04:30 & 05:25 & 94$^{B}$  & 12/11 & 
		09:15 & 09:24 \\ \cline{9-12}
		14 & 28/08 & 16:45 & 17:24 & 41 & 08/09 & 10:30 & 11:00 & 68 & 28/10 & 11:00 & 11:24 & 95  & 12/11 & 11:45 & 
		12:24 \\ \clineB{1-4}{3.5}
		15 & 30/08 & 04:15 & 05:12 & 42 & 08/09 & 23:45 & 00:24 & 69 & 28/10 & 16:00 & 17:24 & 96$^{C}$  & 12/11 & 
		13:45 & 14:08 \\
		16 & 30/08 & 07:30 & 08:36 & 43 & 09/09 & 06:00 & 06:48 & 70 & 29/10 & 12:00 & 04:39 & 97  & 13/11 & 02:00 & 02:54 \\
		17 & 30/08 & 14:20 & 15:36 & 44 & 15/09 & 14:10 & 14:39 & 71 & 29/10 & 12:00 & 18:24 & 98  & 13/11 & 09:30 & 11:08 \\
		18 & 30/08 & 17:30 & 18:36 & 45 & 16/09 & 09:00 & 09:39 & 72 & 31/10 & 10:00 & 11:24 & 99$^{C}$  & 13/11 & 
		11:45 & 12:24 \\
		19 & 30/08 & 19:20 & 19:46 & 46$^{B}$ & 17/09 & 00:15 & 00:39 & 73 & 02/11 & 09:50 & 10:12 & 100 & 13/11 & 
		14:15 & 14:38 \\
		20 & 30/08 & 22:10 & 23:05 & 47$^{B}$ & 17/09 & 03:45 & 04:54 & 74 & 03/11 & 02:00 & 02:36 & 101$^{C}$ & 14/11 
		& 
		00:00 & 00:24 \\
		21 & 31/08 & 00:20 & 01:24 & 48 & 08/10 & 09:10 & 10:24 & 75$^{C}$ & 03/11 & 06:05 & 06:24 & 102 & 14/11 & 
		03:15 & 04:24 \\
		22 & 31/08 & 02:30 & 03:24 & 49 & 10/10 & 20:03 & 22:24 & 76 & 03/11 & 09:10 & 09:24 & 103 & 14/11 & 17:30 & 
		18:24 \\ 
		23 & 31/08 & 05:45 & 06:24 & 50 & 11/10 & 09:00 & 09:44 & 77$^{C}$ & 03/11 & 12:25 & 12:36 & 104$^{B}$ & 15/11 
		& 
		14:45 & 15:24 \\ \cline{13-16}
		24 & 31/08 & 12:00 & 14:12 & 51 & 15/10 & 09:45 & 11:24 & 78$^{M}$ & 05/11 & 14:30 & 15:12 & 105 & 17/11 & 
		02:30 & 02:47 \\
		25 & 31/08 & 19:10 & 20:24 & 52$^{M}$ & 16/10 & 19:30 & 20:24 & 79 & 06/11 & 01:25 & 03:12 & 106$^{B}$& 
		17/11 & 08:15 
		& 
		08:23 \\\clineB{5-8}{3}
		26 & 31/08 & 20:50 & 21:14 & 53 & 17/10 & 04:00 & 05:00 & 80$^{M}$ & 06/11 & 15:40 & 16:12 & 107 & 17/11 & 
		14:25 & 15:11 \\ \clineB{9-12}{3}
		27 & 01/09 & 06:30 & 07:12 & 54$^{C}$ & 17/10 & 09:00 & 09:36 & 81$^{B}$ & 10/11 & 14:20 & 14:54 & 108 & 17/11 
		& 
		18:20 & 
		18:35 \\ \hline	
	\end{tabular}
	}
	\end{center}
\end{table}
\end{landscape}

 We note that, for the computation of the AW, mass and linear speed we used the coronagraphic images where the 
CME propagation direction was closer to the corresponding plane of the sky, i.e. up to 45\,$^\circ$, to reduce 
projection effects. The latter 
cannot be ruled out unless a tridimensional model or tomographic reconstruction of the CME is applied, see e.g. 
\cite{Pluta2019}. However, projection errors effects are milder in our statistical approach due to the different 
propagation directions of the large number of CMEs analyzed.

\subsection{X-ray flaring productivity}
\label{sect:dataFlares}

We collected the AR production of X-ray flares in the 1-8\,$\AA$ band by inspecting 
the catalog of the GOES satellite, available at 
\url{https://www.ngdc.noaa.gov/stp/solar/solarflares.html}.
We included all flares cataloged B-class (10$^{-7}$-10$^{-6}$\,$W\,m^{-2}$) or superior, originating from ARs NOAA 
11089, 11100, 11106, 11112, 11121 or 
11123; and registered from 19 July to 17 November 2010. A total of 162 flares where found, with 127 (78$\%$), 31 (20$\%$) 
and 4 ($2\%$) being of class B, C 
and M, respectively.
	
\section{Results}
\label{sec:results}

\begin{figure}[tp]
	\centering
	\begin{subfigure}{} 
		\includegraphics[width=1.6in, angle=90]{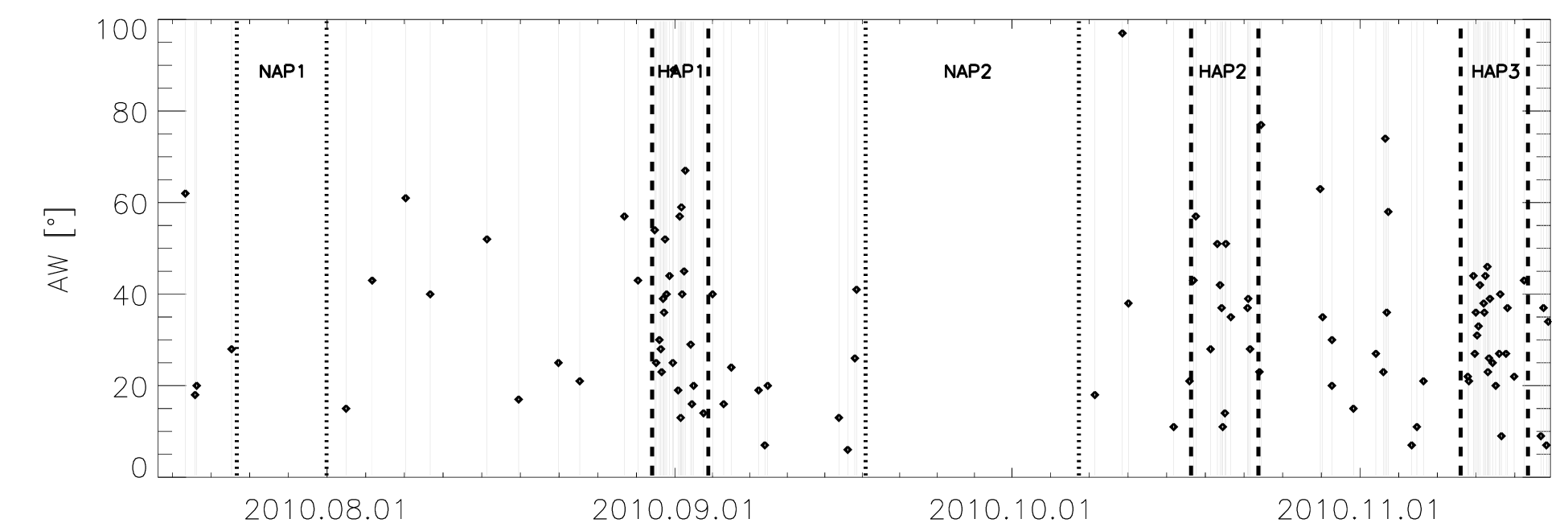}
	\end{subfigure}
	\vspace{-0.3cm}
	\centering
	\begin{subfigure}{}
		\includegraphics[width=1.6in, angle=90]{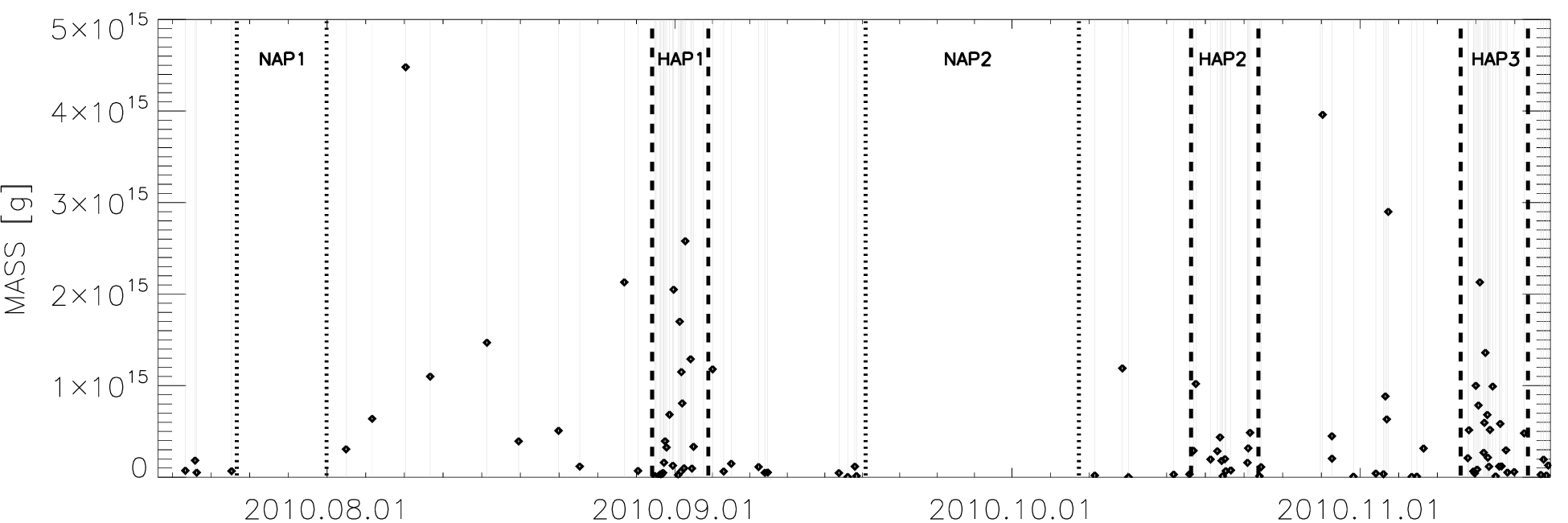}
	\end{subfigure}
	\vspace{-0.3cm}
	\centering
	\begin{subfigure}{}
		\includegraphics[width=1.6in, angle=90]{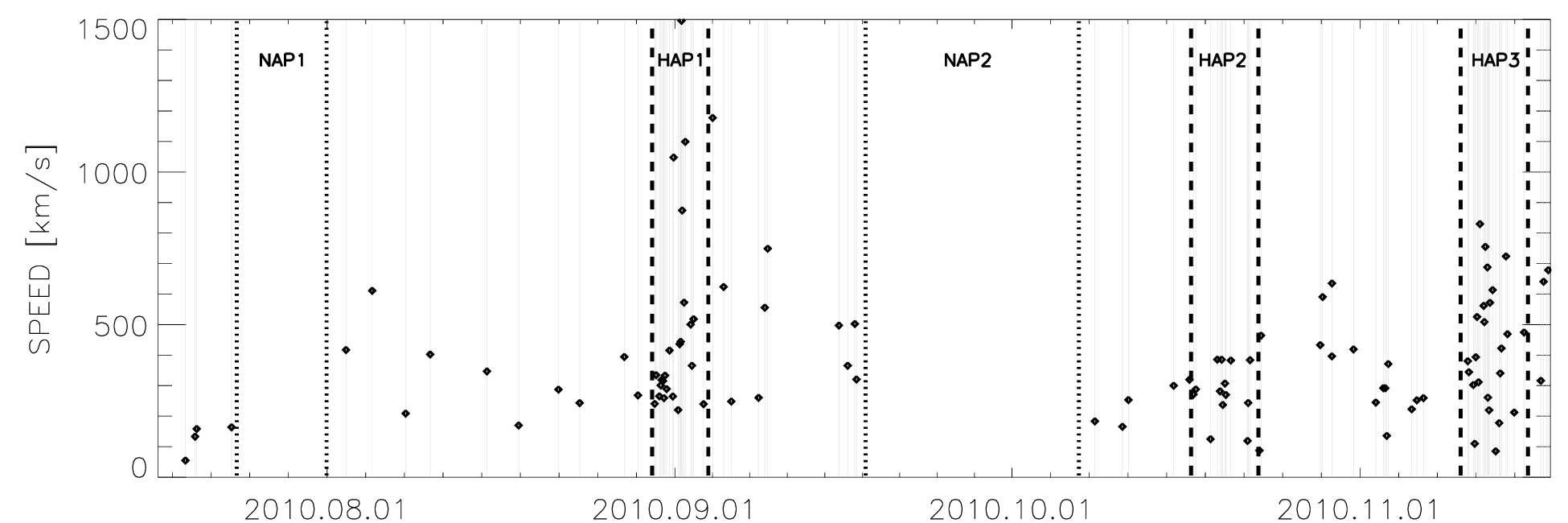}
	\end{subfigure}
	\vspace{-0.3cm}
	\centering
	\begin{subfigure}{}
		\includegraphics[width=1.8in, angle=90]{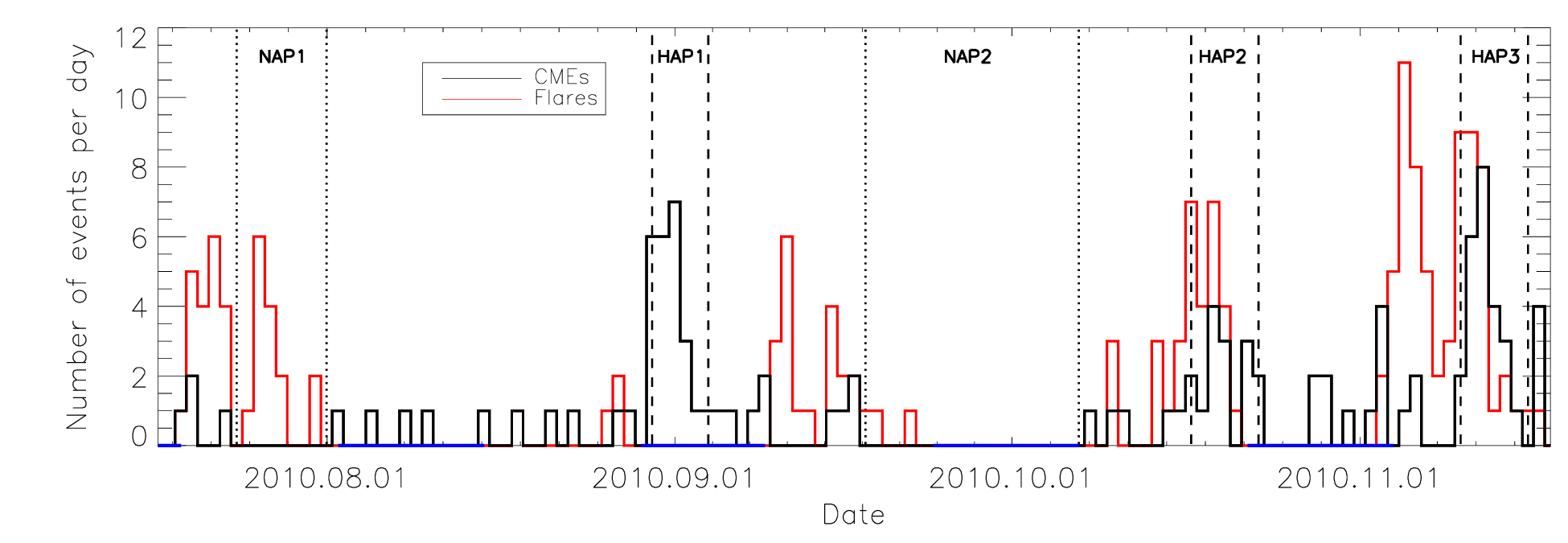}
	\end{subfigure}
	\vspace{-0.4cm}
	\caption{Time evolution of CME properties. From top to bottom we present the measured angular width, mass, linear 
	speed and daily frequency of occurrence (black histogram), for the 108 identified CMEs. The vertical solid grey 
	lines mark the onset 
	of 	a CME. Periods of high (no) CME activity, HAP (NAP), are delimited by the vertical dashed (dotted) 
	lines and 
	labeled in each panel. In the bottom panel, we also indicate the time periods when the AR was in the far side (blue 
	segments in the horizontal axis) and plot the daily frequency of GOES flares (red histogram).}
		\label{fig:propcme}	
\end{figure}

Fig. \ref{fig:propcme} presents the measured properties of the 108 identified ejective events: AW, mass, speed, and 
occurrence rate, as a function of time. The bottom panel (number of CMEs per day) shows periods of time that stand out 
from the 
rest, in that either many CMEs occur in only few days, or no mass ejections are registered during one or more weeks. We 
identified five such time intervals, with two of them being no-activity periods (NAP1 and NAP2) and three of them high 
activity periods (HAP1, HAP2 and HAP3), see labels and vertical dotted and dashed lines in Fig. \ref{fig:propcme}. 
Table \ref{tab:ap} details the time intervals and relevant properties of these activity periods. Note that 
HAP1, HAP2 
and HAP3 combined cover only 16\% (20 days) of the AR lifetime ($\approx124$ days) but account for 59\% (64) of the 
total 
number of produced CMEs. On the other hand, NAP1 and NAP2 combined imply 23\% 
(29 days) of the AR lifetime without a single registered CME.

The bottom panel in Fig. \ref{fig:propcme} also presents the daily frequency of the 162 GOES X-ray flares 
associated to 
the AR (red histogram). Note that, flaring activity is not available when the AR is in the far side (see the blue 
segments in the horizontal axis). A total of 62 CMEs were ejected during AR near-side transits; out of these,  
we could associate 33 to a GOES flare (13, 17 and 3 of class B, C and M, respectively), see Table 
\ref{tab:cme}. This association rate implies that only 53$\%$ (33 out of 62) of the CMEs were accompanied by a flare, 
while 20$\%$ (33 out of 162) of the flares were related to a CME. Moreover, the more energetic the flares, the higher 
is the association rate with CMEs, i.e. 10$\%$ (13 out of 127) , 55$\%$ (17 out of 31), 75$\%$ (3 out of 4) of class B, 
C and M, respectively.
	
For 96 out of the 108 CMEs, we could identify the location of their associated EUV brightening. These are 
	indicated by the 
	blue arrows over the central-meridian magnetograms of each CR in Fig. \ref{fig:armag}. It can be seen that the 
	brightenings are occurring in various places above the AR throughout its lifetime. During HAP1 (CR 2100), the 
	brightening 
	clump 
	together either near the PIL or a small ($\approx5^\circ$ radius) cluster in the positive 
	polarity 
	section. During HAP3 (CR 2103) most of the brightenings (27 out of 36) originated from AR 11123.

\begin{table}[tp]
	\begin{center}
		\caption{Five CME activity periods of the AR under study. We detail various properties (Col. 1) for two no 
		ejective activity periods, NAP1 and NAP2 (Cols. 2 and 4, respectively), three periods of high CME 
		production, HAP1, HAP2 and HAP3 (Cols. 3, 5 and 6, respectively), and the total AR lifetime (Col. 7). 
		The 
		asterisk in the number of X-ray 
		flares means that no data is available because the AR was in the far-side. For each CME property we show the 
		temporal 
		average and standard error. See Fig. \ref{fig:propcme} and the text for extra details.}
		\label{tab:ap}
		{\footnotesize
			\vspace{0.3cm}			
			\begin{tabular}{l|c|c|c|c|c|c}\hline 
				{\bf Property} & {\bf NAP1} & {\bf HAP1} & {\bf NAP2} & {\bf HAP2} & {\bf HAP3} & {\bf Total}\\ \hline
				Starting date  & 24/07 & 30/08 & 18/09 & 17/10 & 10/11 & 19/07 \\ 
				Ending date   & 01/08 & 04/09 & 07/10 & 23/10 & 16/11 & 17/11\\ 
				Duration [days] & 9 & 6 &20 & 7 & 7 & 124 \\ 
				CMEs & 0 & 24 & 0 & 15 & 24 & 108 \\ 
				X-ray flares & 15 & * & * & 23 & 31 & 162\\ 
				Asoc. CME-flare & 0 & * & * & 6 & 17 & 33\\ 
				CME AW [$^\circ$]& - & 38$\pm$19 & - & 38$\pm$16 & 32$\pm$10 & 33$\pm$18  \\ 
			    CME Mass [$10^{14}$g]& - & 5.7$\pm$7.4 & - & 2.6$\pm$2.5 & 4.7$\pm$5.1 & 5.6$\pm$10.5 \\ 
			    CME Speed [km s$^{-1}$] & - & 514$\pm$353 & - & 282$\pm$110 & 429$\pm$205 & 394$\pm$235 \\ \hline
			\end{tabular}
		}
	\end{center}
\end{table}

\begin{center}
	\begin{figure}[tp]
		\centering
		\begin{subfigure}{}	
			\includegraphics[width=2in,angle=90]{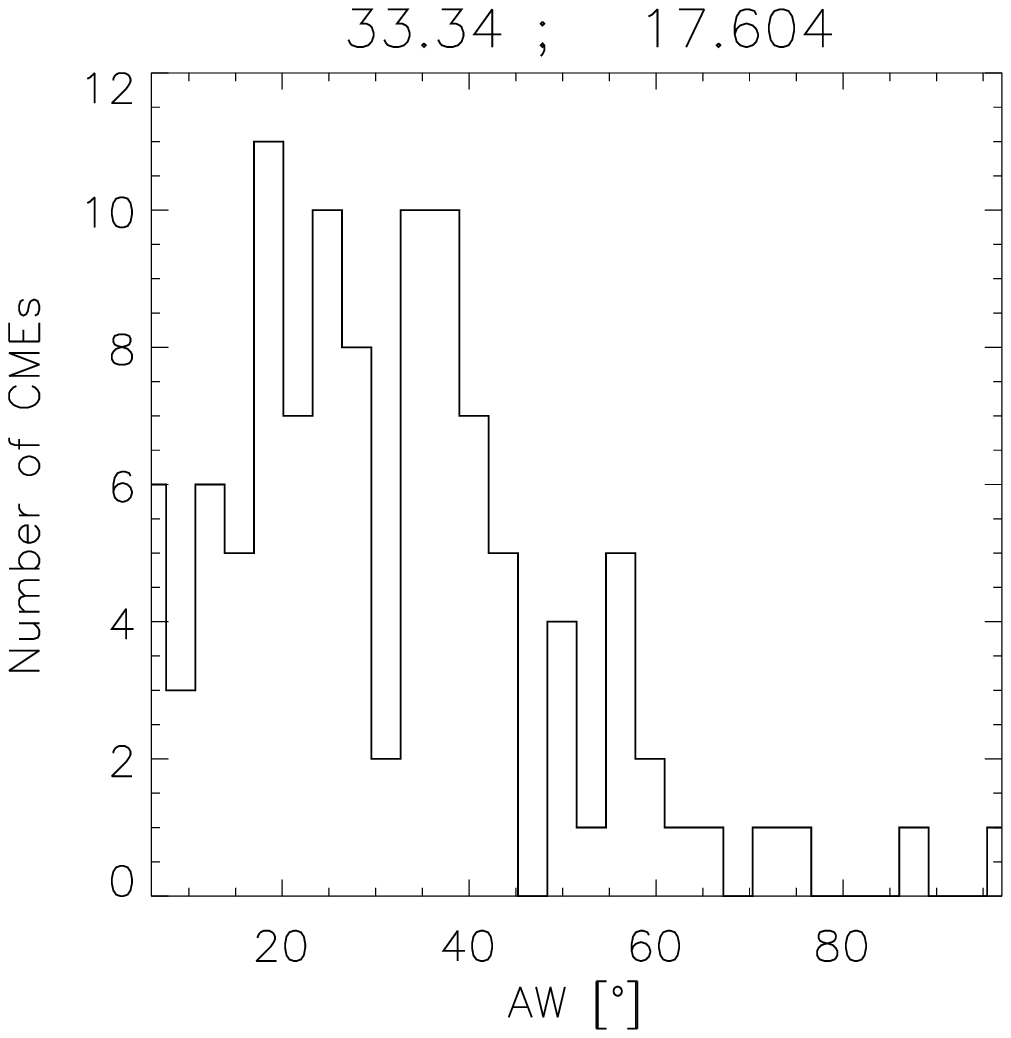} 
		\end{subfigure}
		\begin{subfigure}{}
			\includegraphics[width=2in, angle=90]{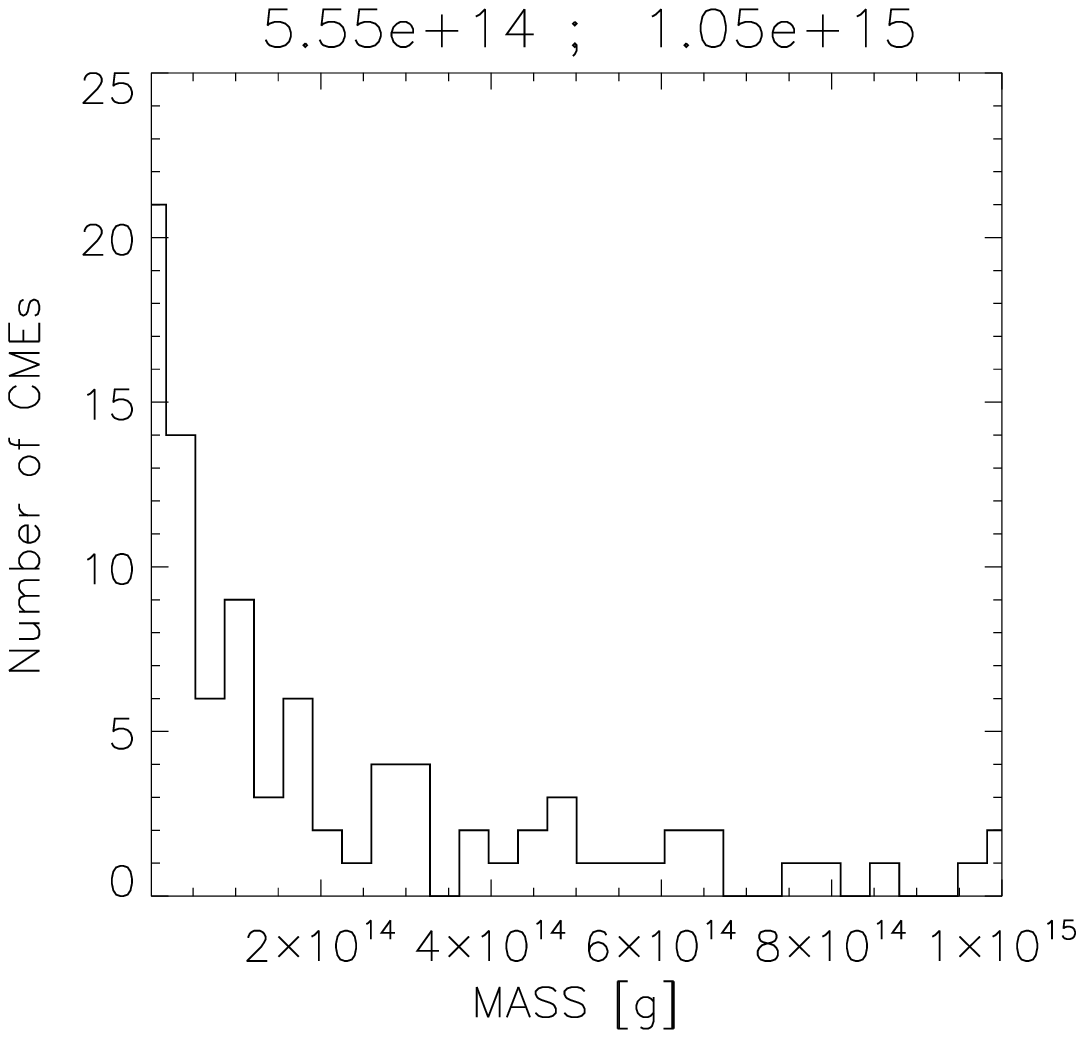}
		\end{subfigure}
		\begin{subfigure}{}
			\includegraphics[width=2in, angle=90]{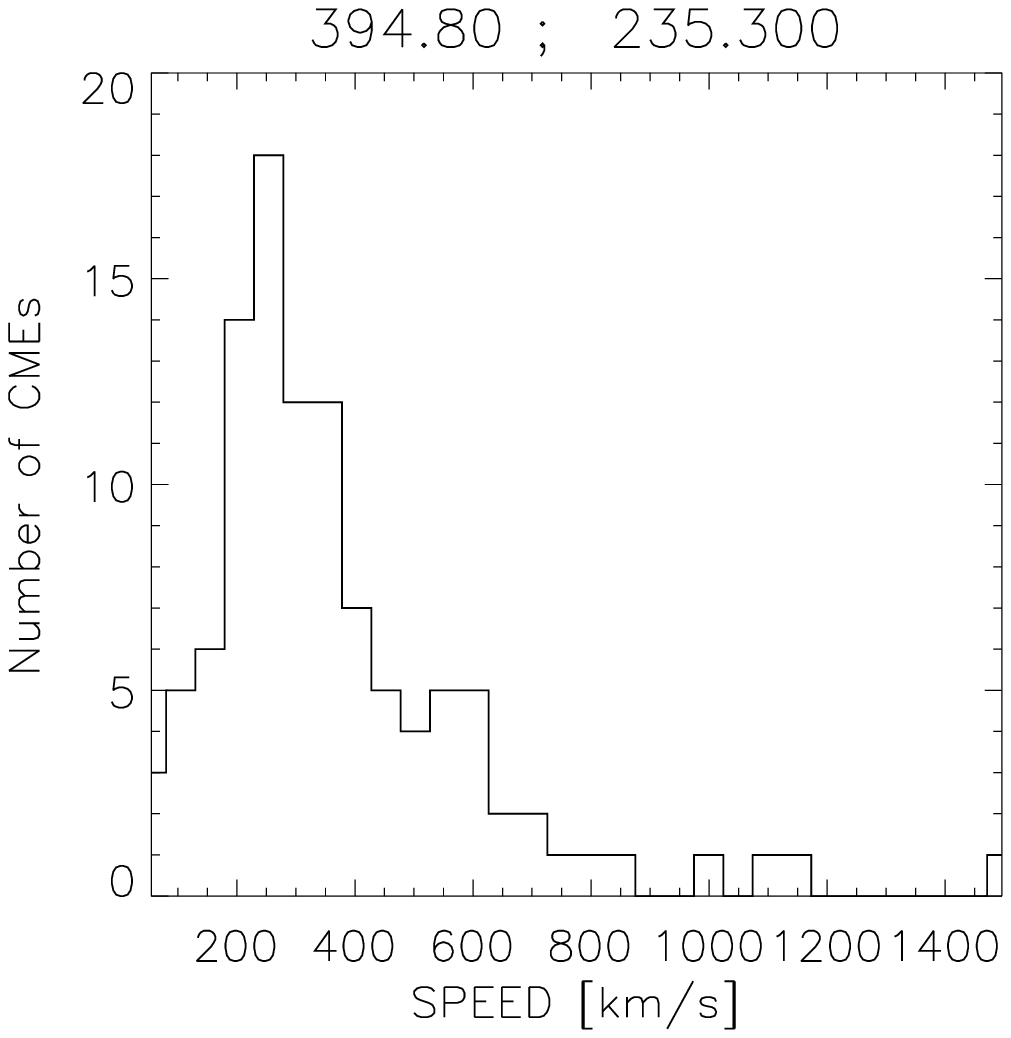}
		\end{subfigure}
		\begin{subfigure}{}
			\includegraphics[width=2in, angle=90]{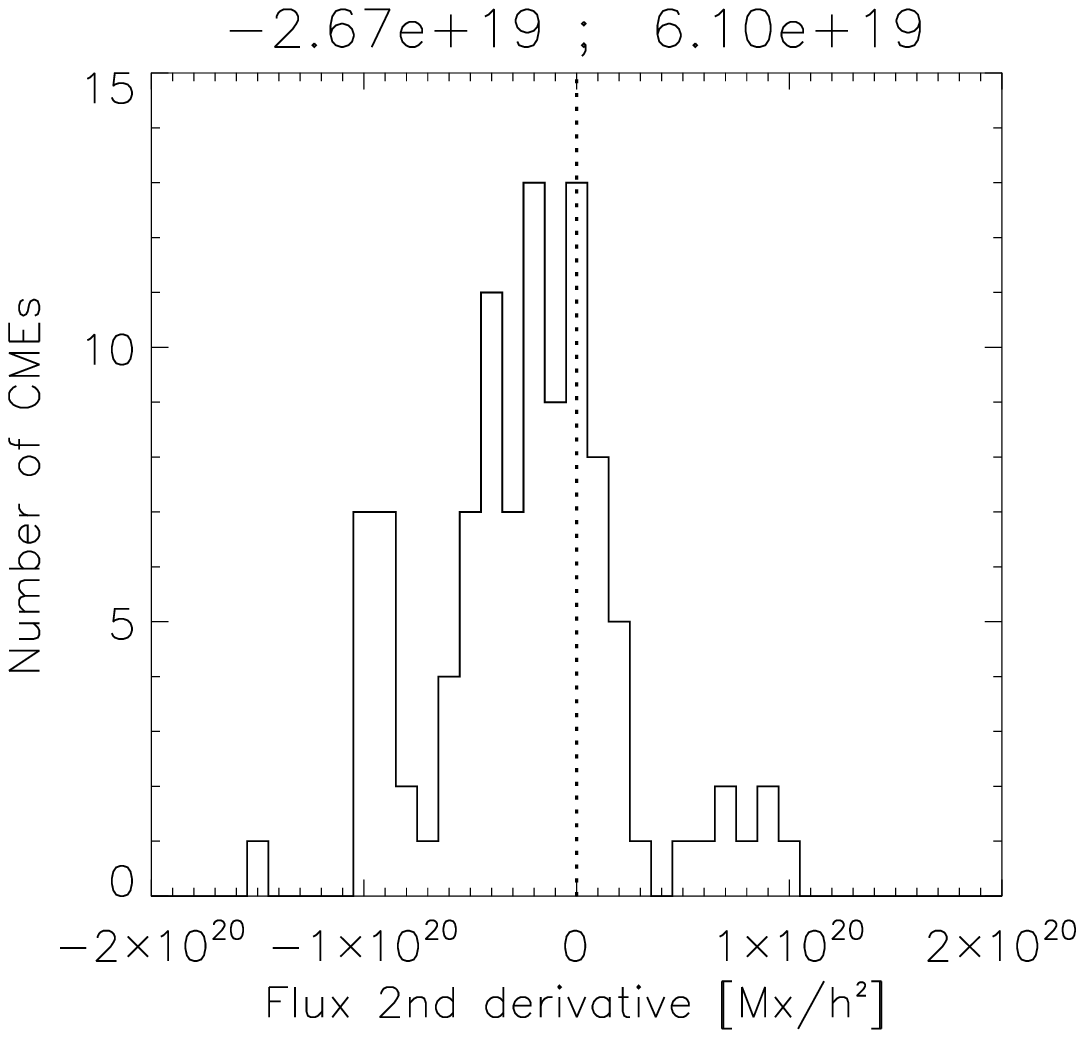}
		\end{subfigure}
		\caption{Distribution of the AW, mass, speed and value of the magnetic flux second derivative during onset (see axes labels) for the 108 identified CMEs. The title in each panel presents the average and standard error, 		respectively. Further details are described in the text.}
		\label{fig:hist}
	\end{figure}
\end{center}

From the top panel in Fig. \ref{fig:propcme} and Table \ref{tab:ap}, it can be seen that the AW distribution of HAP3 is 
narrower than those for HAP1 and HAP2, with all CMEs having AWs below $\approx45^\circ$. The largest AW 
(97$^\circ$) 
was 
detected out of the high activity periods for a rather isolated CME occurring on 10 October at 20:03. Fig. 
\ref{fig:hist} presents the distribution of all AWs, including their average ($\approx33^\circ$) and standard error 
($\approx18^\circ$).  On the other hand, the distribution of all CME masses is strongly skewed (4.09) towards high 
values, see Fig. \ref{fig:hist}, with a mean of $5.6\times10^{14}$\,g which is slightly higher to the value reported by 
\cite{Vourlidas2010b}, $3.9\times10^{14}$\,g. The dispersion and mean 
value of the masses during HAP2 are 
approximately half of the values registered for HAP1, HAP3 and the full set, see Table \ref{tab:ap}. Regarding speeds, the 
largest were registered during HAP1, with 4 events having speeds above 1000\,km\,s$^{-1}$. The dispersion and mean 
value of the speeds during HAP2 are also below the figures registered for HAP1, HAP3 and the full set, see Table 
\ref{tab:ap}. The overall speed distribution is mildly skewed towards high values (1.83) with an average 
($\approx394$\,km\,s$^{-1}$) within the slow solar wind speed range ($<$500\,km\,s$^{-1}$, \citealt{Abbo2016}), see 
Fig. 
\ref{fig:hist}.

\begin{figure}[tp]
	\centering 
	\begin{subfigure}{}	
		\includegraphics[width=1.6in, angle=90]{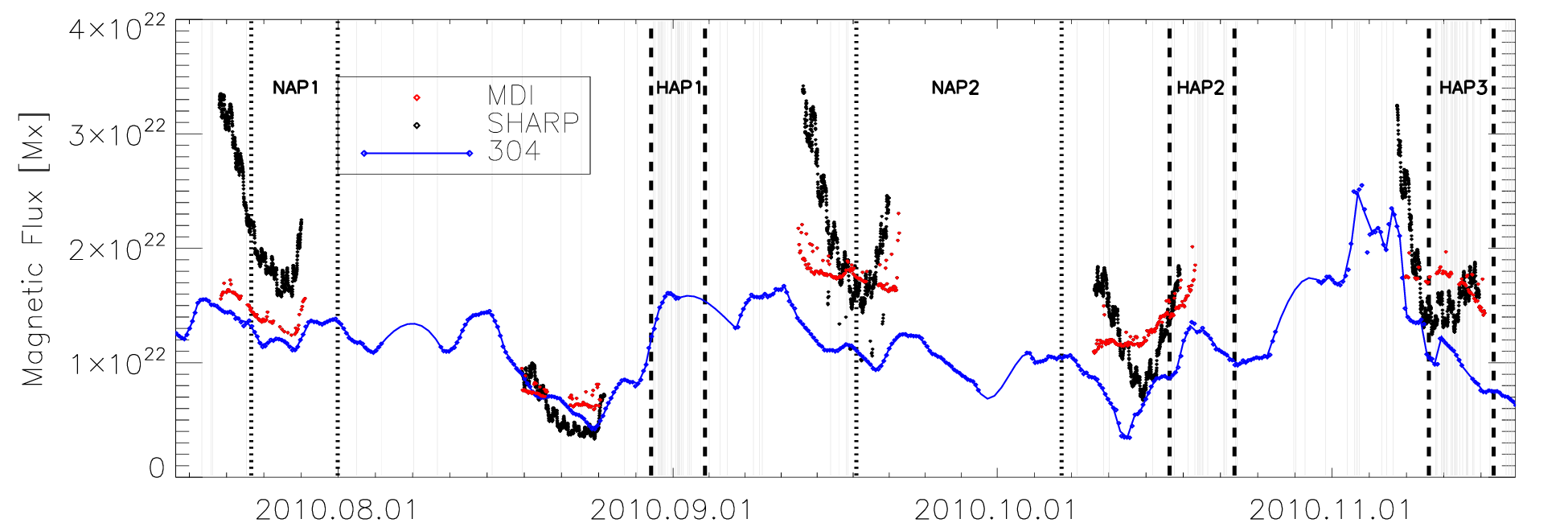}
	\end{subfigure}
	\vspace*{-0.3cm}
	\centering 
	\begin{subfigure}{}
		\includegraphics[width=1.6in, angle=90]{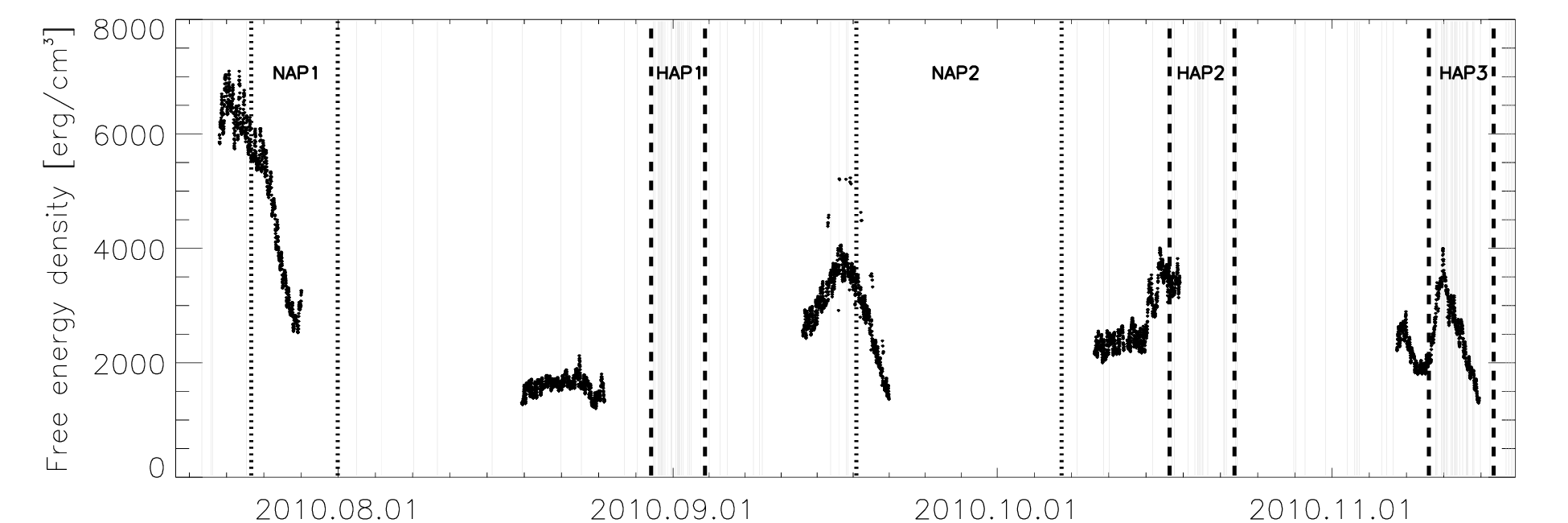}
	\end{subfigure}
	\vspace*{-0.3cm}
	\centering 
	\begin{subfigure}{}
		\includegraphics[width=1.6in, angle=90]{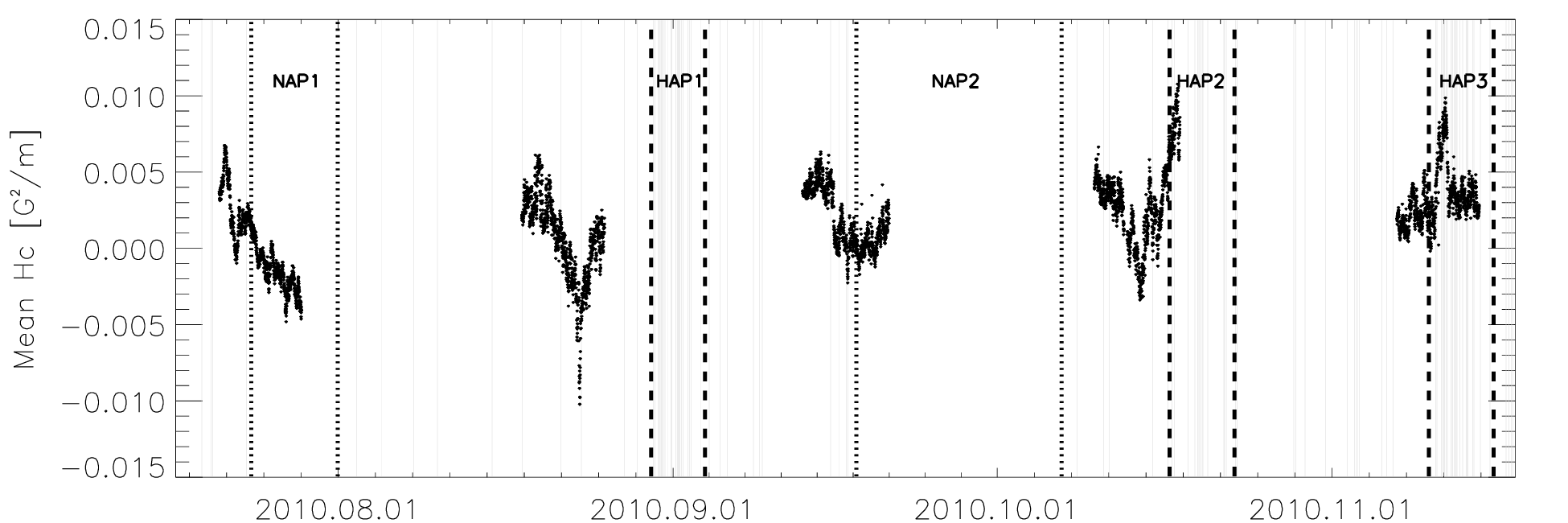}
	\end{subfigure}
	\vspace*{-0.3cm}
	\centering 
	\begin{subfigure}{}
		\includegraphics[width=1.8in, angle=90]{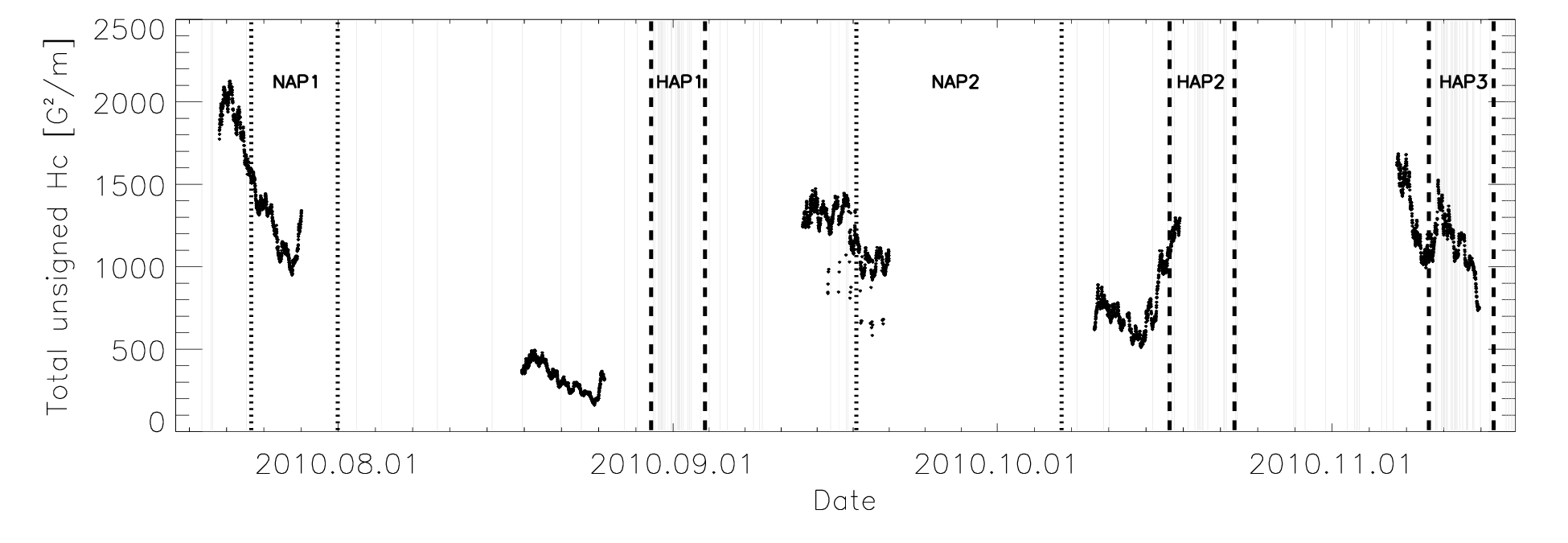}
	\end{subfigure}
	\vspace*{-0.4cm}
	\caption{Time evolution of SHARP photospheric magnetic field parameters. From top to bottom we show the magnetic 
	flux estimated via three different sources (see legend and Sect. \ref{sect:dataAR}), a proxy for the free magnetic 
	energy density,  the mean current helicity $Hc$ and the total unsigned $Hc$. The vertical lines have the same 
	meaning 
	than in Fig. \ref{fig:propcme}.}
	\label{fig:propar}
\end{figure}

Fig. \ref{fig:propar} shows the AR photospheric field properties for the full analyzed time interval.  As explained in 
Sect. \ref{sect:dataAR}, HMI magnetograms and thus SHARP data are only available during the near-side passages of the 
AR, while HAP1 occurred when the AR was on the far side. The total unsigned flux curves derived from three 
different sources are plotted in the top panel of Fig. \ref{fig:propar} with different colors. They differ from each 
other due to the limitations of each technique, which have been addressed in Sect. \ref{sect:dataAR}. In particular, 
the flux derived from the 304\,\AA~images underestimates the SHARP flux during the first and third near-side passages. 
However, the general trend of variation is similar in all three flux estimations, except during the end of the third 
and fifth near-side passages. The flux derived from the 304\,\AA~images has been interpolated and 
smoothed with a 24\,h window, to reduce the influence of flaring activity (see \citealt{ugarteurra2015}), and fitted 
with an spline to obtain an continuous estimation. 

\section{Discussion and conclusions}
\label{sec:discussion}
To provide further insights on the ejective activity variation of ARs, we have made use of the advantageous positioning 
of the STEREO and SOHO spacecraft to track and characterize the white-light ejecta of a long-duration AR during its 
complete lifetime (spanning approximately 124 days). We found a large number of mass ejections (108), distributed 
non-uniformly in time, i.e. 56$\%$ of all CMEs occurred in 16$\%$ of the AR lifetime.  Motivated by the large number 
and clustering of the CMEs within the studied time interval, we focus the discussion below on comparing the occurrence 
of full ejective activity periods with the GOES X-ray flaring activity, and the long-term (multi-day, see 
\citealt{Green2018}) temporal variation of the AR photospheric properties using SDO and STEREO. Such analysis differs 
from the more frequent studies of the short-term (tens of hours) evolution of the AR magnetic properties, preceding the 
occurrence of single (or few) CMEs and/or flares (see Sect. \ref{sec:intro}).

Fig. \ref{fig:propar} shows that HAP1 and HAP2 take place during or after periods of persistent ($\approx$3 days) 
magnetic 
flux 
increments. 
Using the 304\,\AA~proxy these increments are of $\approx1.2\times10^{22}$ and $\approx1\times10^{22}$\,Mx for HAP1 and 
HAP2, respectively. Moreover, it stands out that a large portion of the CMEs were ejected when the flux variation was 
decreasing, i.e. reaching a plateau. The bottom-right histogram presented in Fig. 
\ref{fig:hist} quantifies this, showing that 73 out of 108 CMEs occur when the second temporal derivative of the 
magnetic flux is negative. Note that the number of events analyzed is small and only a proxy of the magnetic 
flux is used, thus a hard conclusion from these results cannot be drawn. However, many other single or few-events studies 
report 
CMEs occurring after the flux emergence, e.g. 
\citealt{romano2014, bobra2014,li2012,Mandrini-etal2014,Jiang2014}. This is likely related to the fact that the 
non-potential energy may continue 
increasing during the flux stabilization period, as the shearing photospheric motions twist the emerged flux rope 
building up the coronal field helicity. During HAP2, the flux emergence is also 
followed by a clear increase of the field non-potentiality before 
the burst of several CMEs, suggesting again the twisted flux emergence and its accompanying shearing photospheric 
motions as the main free energy contributors. This is 
manifested by the jumps in the magnetic free energy density 
($\approx2000$\,erg\,cm$^{-3}$), mean $Hc$ ($\approx0.012$\,$G^2$m$^{-1}$) and total unsigned $Hc$ 
($\approx700\,$G$^2$m$^{-1}$), see Fig. \ref{fig:propar}. 

The case of HAP3 is different in that it starts 
during the decay phase of AR 11121. A persistent decrease ($\approx$5 days) of 
the total unsigned magnetic flux is registered before HAP3 begins (according to the SHARP data and the 
304\,\AA~proxy) and continues with only moderate increases and recurrent dips 
($<5\times10^{21}$\,Mx, according to the SHARP data) until the end of the period. During the decay phase of 
bipolar ARs,
the CME activity can increase due to the flux cancellations produced at the PIL or by the motion of magnetic elements \citep{vandriel2015}, which induce the successive formation and 
eruption of flux ropes (because it favors the onset of torus 
instabilities \citealt{forbes1991}, among others). However, for the case of HAP3, the persistent increase 
in the free magnetic energy density and $Hc$ (for a $\approx$3 days period and similar in value to the ones registered 
before HAP2, see Fig. 
\ref{fig:propar}) are most likely related to the emergence of AR 11123, from 
which most of the EUV brightenings in HAP3 originate (see Fig. \ref{fig:armag}). As detailed by 
\cite{Mandrini-etal2014}, a series 
of fast bipolar emergences 
formed AR 11123 between 9 and 10 November. This produced an increase in AR 11123 flux (partially masked in Fig 
\ref{fig:propar} because we consider the full AR complex including AR 11121, see Fig. 5 in \citealt{Mandrini-etal2014}) 
due to the creation and subsequent topological evolution of the magnetic field in the following two days. After this 
period, the free energy density and the mean $Hc$ peak, suggesting that the non-potential energy was built by 
photospheric shearing motions, and the most active portion of HAP3 begins. Note that, during most of HAP3 the free 
magnetic 
energy density reduces monotonically due to the successive CMEs.

We note that, even though HAP1 and HAP2 were preceded by a flux increment and occurred when the AR was a simple bipole, 
the aggregate CME properties reported in Table \ref{tab:ap} do not substantially differ from those of HAP3, which 
occurred after a strong flux decrease when AR11123 emerged and the photospheric field was a more complex quadrupole 
(two bipoles).

Substantial X-ray flaring activity was registered during HAP2 and HAP3 (54 flares), however only 40$\%$ (6 out of 15) 
of CMEs in 
HAP2  were associated to a flare. This association increases to 71$\%$ (17 out of 24) for HAP3, which 
includes CMEs that are on average 52$\%$ faster and 81$\%$ more massive, see Table \ref{tab:ap}. This is in agreement 
with 
the well known fact that more energetic CMEs tend to be preceded by bigger flares, see e.g. \cite{Webb2012} and references 
therein. On the other hand, 98$\%$ of the flares registered during the AR lifetime are of B or C class, and most of the CMEs are slow (75$\%$ having speeds $<$600\,$kms^{-1}$). Moreover, no X-class flare was associated to the AR, although the fastest 4 CMEs, within the speed range commonly 
associated to X-class flares 
\citep[$\approx$1500 km s$^{-1}$, see][]{Yashiro2005}, were produced during the back-side transit in HAP1, not visible 
by GOES.

The two main periods of no ejective activity, NAP1 and NAP2, begin during a phase of strong reduction of the 
free magnetic energy density, although the initial values are comparable to those found at the beginning of HAP2 and 
HAP3. NAP1 also starts during a period of $Hc$ and flux reduction, suggesting that the non-potential energy was reduced 
by flux cancellation, likely related to the decay of the small western bipole present during the first 
rotation of AR11089 (see top-left panel in Fig. \ref{fig:armag}). Such a flux cancellation is accompanied by flaring 
reconnections (see Table \ref{tab:ap}), however, no CMEs are observed.  This can be due to the fact that the AR is 
young 
and thus no well formed flux ropes are present and/or due to a more efficient confinement by the overlying strapping 
field, see e.g. \cite{romano2014}. On the other hand, NAP2 begins with a flux rise (according to 
SHARP data) and not a marked $Hc$ increment, suggesting a mechanism other than flux cancellation, diffusion, or field 
ejection to explain the free energy reduction, e.g. flux emergence with an helicity sign opposite to the predominant. 
The main results of the long-term, multi-viewpoint study reported here are summarized below:

\begin{itemize}{
\item  56$\%$ of the 108 CMEs identified occur in three activity periods (HAP1, HAP2 and HAP3) spanning 16$\%$ of the 
AR 
life time($\approx$ 124 days).	Two periods of no CME activity where identified (NAP1 and NAP2) spanning 23$\%$ of the 
AR life time.
\item HAP1 and HAP2 take place after periods of persistent ($\approx$3 days) magnetic flux increment ($\approx 
1\times10^{22}$\,Mx) and free magnetic energy (only measurable for HAP2).
\item 73 out of 108 CMEs occur when the flux change rate is decreasing, i.e. during intervals of negative second time 
derivative.
\item HAP3 occurs during the decay phase of the AR 11123 and a persistent reduction ($\approx$ 5 days) of the magnetic 
flux. The high CME activity is related to the free energy increment produced by the flux injection and photospheric 
motions induced by the emergence of AR 11121.
\item There is no statistical difference among the aggregate CME properties of the three activity periods.
\item 62 CMEs occurred during front-side transits, 33 where associated to one of the 162 GOES flares identified. 
The more energetic the flare, the higher the association rate, i.e. 10$\%$ (13 out of 127) , 55$\%$ (17 out of 31), 
75$\%$ (3 out of 4) of class B, C and M, respectively.
\item Most of the CMEs found are slow (75 $\%$ having speeds $<600\,km^{-1}$) and thus 98$\%$ of the flares are of B 
and C class.
\item NAP1 and NAP2 occur during phases of strong reduction of free magnetic energy. NAP1 occur during a flux reduction 
interval accompanied by flaring activity but no CMEs, likely because the AR was young and no sizable filamentary 
structure was present.
\item No obvious correlation was found between the long-term variation of the average photospheric 
properties, and the values of the aggregated CME characteristics of the high activity periods.
}
\end{itemize}

Regarding the last point, it is known that 
the photospheric field properties are only a partial indication of the likelihood of an AR to produce a mass ejection, 
see e.g. \cite{Green2018, Mandrini2014b, romano2014}. The characteristics of the higher coronal field above the 
potential CME source region, e.g. the presence of streamers, are not addressed here and strongly affect the CME 
production and early kinematics. An additional effect, that we plan to further study, is related to the fact that we 
employ photospheric quantities averaged over the full SHARP patch, while the EUV brightenings associated to the CMEs 
tend to cluster in specific sectors within the AR.
\\
\\
\textbf{Acknowledgements:} HC, CHM and MLF are members of the Carrera del Investigador Cient{\'i}fico (CONICET). FAI 
and FML are fellows of CONICET. LM acknowledges a scholarship from UTN and from the CIN. The authors appreciate 
financial support from the Argentine grants PICT 2012-973 (ANPCyT) and PIP 2012-01-403 (CONICET). FAI, HC, LM, and FML 
thank support from project UTN UTI4035TC and UTI4915TC. IUU acknowledges support by a grant from the  NASA Heliophysics 
Guest 
Investigator program (NNH16AC71I). The authors acknowledge the use of data from the STEREO (NASA), SDO (NASA), and SOHO 
(ESA/NASA) missions.  These data are produced by the AIA, HMI, SECCHI, LASCO, and MDI international consortia.

\bibliography{../../iglesias}  

\end{document}